\documentclass[aps,prr,superscriptaddress,twocolumn,floatfix,a4paper]{revtex4-2}

\usepackage{graphicx,graphics,epsfig}   
\usepackage{dcolumn}    
\usepackage{bm}         
\usepackage{amsmath}    
\usepackage{verbatim}   
\usepackage{color}      
\usepackage{subfigure}  
\usepackage{times,natbib}
\usepackage{amsmath,amsfonts,amssymb,graphics,graphics,color,times}

\usepackage[cp1250]{inputenc}
\usepackage[T1]{fontenc} 

\usepackage{xcolor} 
\usepackage{url}

\usepackage[normalem]{ulem}

\usepackage{orcidlink} 

\usepackage{hyperref}
\hypersetup{colorlinks=true, citecolor=blue, urlcolor=blue, linkcolor=blue}

\usepackage{latexsym}
\usepackage{amsmath}
\usepackage{amssymb}
\usepackage{amsfonts}
\usepackage{amsthm}
\usepackage{mathrsfs}
\usepackage{color,verbatim,graphics}
\usepackage{psfrag}
\DeclareMathAlphabet{\mathrsfs}{U}{rsfs}{m}{n}
\DeclareMathAlphabet{\mathpzc}{OT1}{pzc}{m}{it}
\DeclareMathAlphabet{\matheus}{U}{eus}{m}{n}
\DeclareMathAlphabet{\mathbbold}{U}{bbold}{m}{n}

\def\one{\leavevmode\hbox{\small1\normalsize\kern-.33em1}}

\newcommand{\CC}{\mathbb{C}}

\renewcommand{\qed}{\ensuremath{\hfill \blacksquare}}

\newcommand{\ba}{\begin{eqnarray}}
\newcommand{\ea}{\end{eqnarray}}
\newcommand{\ban}{\begin{eqnarray*}}
\newcommand{\ean}{\end{eqnarray*}}

\newcommand{\ket}[1]{|#1\rangle}
\newcommand{\bra}[1]{\langle#1|}

\newcommand{\ketbra}[2]{|#1\rangle\langle#2|}

\newcommand{\rhopptfirst}{{\varrho_{\rm F1}}}
\newcommand{\rhopptsecond}{{\varrho_{\rm F2}}}
\newcommand{\rhopptn}{{\varrho_{{\rm F}n}}}
\newcommand{\rhopptfirstp}{{\varrho_{{\rm F}1}^{(p)}}}

\newcommand{\rhopptfnp}{{\varrho_{{\rm F}n}^{(p)}}}
\newcommand{\rhopptfirstr}{\varrho_{{\rm F}1}^{(1-r)}}
\newcommand{\Hop}{{\mathcal H}}

\begin{document}

\title{Bound entangled singlet-like states for quantum metrology}

\author{K\'aroly F. P\'al}
\email{kfpal@atomki.hu}
\affiliation{Institute for Nuclear Research, P. O. Box 51, H-4001 Debrecen, Hungary} 

\author{G\'eza T\'oth\,\orcidlink{0000-0002-9602-751X}}
\email{toth@alumni.nd.edu}
\homepage{\\ http://www.gtoth.eu}
\affiliation{Department of Theoretical Physics, University of the Basque Country UPV/EHU, P. O. Box 644, E-48080 Bilbao, Spain}
\affiliation{Donostia International Physics Center (DIPC), P. O. Box 1072, E-20080 San Sebasti\'an, Spain}
\affiliation{IKERBASQUE, Basque Foundation for Science, E-48013 Bilbao, Spain}
\affiliation{Institute for Solid State Physics and Optics, Wigner Research Centre for Physics, P. O. Box 49, H-1525 Budapest, Hungary}

\author{Erika Bene}
\email{bene@atomki.hu}
\affiliation{MTA Atomki Lend\"ulet Quantum Correlations Research Group, Institute for Nuclear Research, P. O. Box 51, H-4001 Debrecen, Hungary} 

\author{Tam\'as V\'ertesi\,\orcidlink{0000-0003-4437-9414}}
\email{tvertesi@atomki.hu}
\affiliation{MTA Atomki Lend\"ulet Quantum Correlations Research Group, Institute for Nuclear Research, P. O. Box 51, H-4001 Debrecen, Hungary}

\date{\today}


\begin{abstract}
Bipartite entangled quantum states with a positive partial transpose (PPT), i.e., PPT entangled states, are usually considered very weakly entangled. 
Since no pure entanglement can be distilled from them, they are also called bound entangled. 
In this paper we present two classes of ($2d\times 2d$)-dimensional PPT entangled states for any $d\ge 2$ which outperform all separable states in metrology significantly. 
We present strong evidence that our states provide the maximal metrological gain achievable by PPT states for a given system size. When the dimension $d$ goes to infinity, the metrological gain of these states becomes maximal and equals the metrological gain of a pair of maximally entangled qubits.
Thus, we argue that our states could be called ``PPT singlets.'' 

\vspace{1em}
\noindent DOI: \href{https://doi.org/10.1103/PhysRevResearch.3.023101}{10.1103/PhysRevResearch.3.023101}
\end{abstract}

\maketitle

\section{Introduction}
\label{sec:intro}

Entanglement is at the core of quantum physics and a useful resource with many fruitful applications in quantum information~\cite{horo_review,GT_review}. With the aid of entanglement, tasks that are otherwise impossible become achievable. Such famous tasks are for instance  quantum teleportation, superdense coding and quantum error correction. However, the important question arises precisely which entangled states are useful in a certain application. 

\begin{figure}[t!]
\includegraphics[width=8cm]{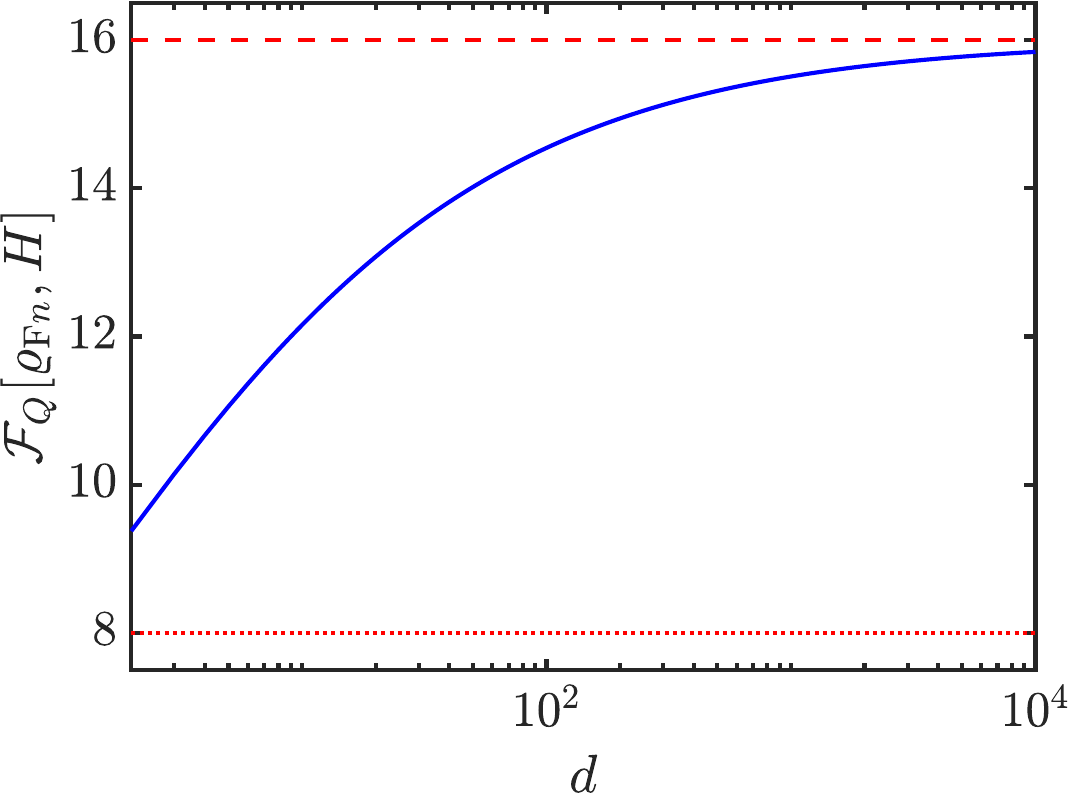}
\caption{Metrological performance of the bound entangled states presented in this paper. (solid) The dependence of the quantum Fisher information ${\cal F}_Q[\rhopptn,H]$ on the dimension $d,$ given also analytically in  Eq.~(\ref{eq:qfinfo}).
 $(2d)\times(2d)$ PPT quantum states are considered here and the Hamiltonian is defined in Eq.~(\ref{eq:Ham}).
Note that the dependence is the same for the two families of states denoted by $\rhopptn$ for $n=1,2.$ 
(dashed) Maximum for the quantum Fisher information for bipartite quantum states.
(dotted) Maximum for the quantum Fisher information for separable quantum states.
Since the states have a quantum Fisher information larger than this value for all $d,$ they are evidently entangled.
} \label{fig:FQd}
\end{figure}

In this paper, we are concerned with quantum states with a positive partial transpose (PPT,~\cite{PPT}), which are considered weakly entangled. We present concrete examples of such states that are useful for metrology.
We apply the quantum Fisher information as a figure of merit of the metrological usefulness of the states, which is defined as \cite{Metrology1,Metrology2,Metrology3}
\begin{equation}
\label{eq:FQ}
{\cal F}_Q[\varrho,\Hop]=2\sum_{\mu,\nu}\frac{(\lambda_{\mu}-\lambda_{\nu})^{2}}{\lambda_{\mu}+\lambda_{\nu}} \left|\langle \mu \vert {\Hop} \vert \nu \rangle \right|^{2},
\end{equation}
where $\mathcal H$ is the Hamiltonian of the system and the eigendecomposition of $\varrho$ is given as
$\varrho=\sum_{\mu}\lambda_{\mu}\ketbra{\mu}{\mu}.$
The larger the quantum Fisher information, the better the quantum state $\varrho$ for metrology for a fixed Hamiltonian $H.$

We now summarize the main result of this paper. 
We will present two families of bipartite quantum states for $(2d)\times (2d)$ systems, denoted by $\rhopptfirst$ and $\rhopptsecond.$

{\bf Observation 1}.---For both families of states, 
\begin{equation}
{\cal F}_Q[\rhopptn,H]=\frac{16\sqrt{d}}{1+\sqrt{d}},
\label{eq:qfinfo}
\end{equation}
holds, see also Fig.~\ref{fig:FQd}. The Hamiltonian corresponding to the $(AA')(BB')$ partition is
\begin{equation}
H=\sigma^z_A\otimes \openone_B\otimes  \openone_{A'B'}+ \openone_A\otimes\sigma^z_B\otimes  \openone_{A'B'}, 
\label{eq:Ham}
\end{equation}
where the dimension of $A'$ and $B'$ is $d.$ Based on Eq.~(\ref{eq:qfinfo}),  we see that the quantum Fisher information given in Eq.~\eqref{eq:qfinfo} approaches 16  for large $d,$ which will be shown to be the maximum achievable by entangled states. Thus, PPT states turn out to be almost as useful as entangled states with a non-positive partial transpose in this metrological task. The proof will be given later in this paper, together with the definition of the quantum states.

We will see that the maximum of ${\cal F}_Q[\varrho,H]$ for separable states is $8.$ Since the quantum Fisher information of the states $\rhopptn$ given in Eq.~(\ref{eq:qfinfo}) is larger than this value for all $d,$ the states $\rhopptn$ are entangled.
(See a confirmation of this fact also in Fig.~\ref{fig:FQd}.)

The starting point of our search for metrologically useful PPT states was the family of such states found numerically for bipartite systems in Ref.~\cite{Toth18}. These states have been obtained from a very efficient numerical maximization of the quantum Fisher information over the set of PPT states, thus we can expect that they have the largest quantum Fisher information among PPT states for the system sizes considered.  These states have been found for bipartite systems up to dimension $12\times12.$ In this paper, we restrict our attention to qudits with an even dimension, since for this case it seems to be easier to look for bound entangled states analytically. Then, we can say that bound entangled states have been found numerically for $2d\times2d$ systems for $d=2,3,4,5,6.$ For the $4\times4$ case (i.e., for $d=2$), even an analytical construction has been presented \cite{Toth18}.

In this paper, we looked for similar families of bound entangled states defined analytically. We now give some more details about the states presented.  The family $\rhopptfirst$ are only the states obtained by Badzi{\k{a}}g {\it et al.}~\cite{Badziag}. The states of this family are invariant under the partial transposition. Their metrological performance is the same as that of the states found numerically \cite{comment}. All these states are states in a  $2d\times2d$ system. 
Their rank is
\begin{align}
{\rm rank}(\rhopptfirst)=d^2+d.\label{eq:rhorank1}
\end{align}
The state is invariant under a  partial transposition, hence
\begin{align}
{\rm rank}(\rhopptfirst^\Gamma)={\rm rank}(\rhopptfirst)=d^2+d,\label{eq:rhorank1pt}
\end{align}
where $\Gamma$ denotes partial transposition according to $BB'.$
For $d=2,$ the density matrix can easily be mapped by trivial unitary operations and relabeling the parties within the two subsystems to the $4\times4$ state mentioned above, given in  Ref.~\cite{Toth18}. 

In this paper, we present another family of states, $\rhopptsecond.$ These states are different from $\rhopptfirst.$  Their rank is larger
\begin{align}
{\rm rank}(\rhopptsecond)=d^2+2d.\label{eq:rhorank2}
\end{align}
They are not invariant under the partial transposition and the rank of their partial transpose is  
\begin{align}
{\rm rank}(\rhopptsecond^\Gamma)=2d^2+d.\label{eq:rhorank2pt}
\end{align}
Their metrological performance is the same as that of the first family for all $d$. In the definition of the $\rhopptsecond$ states a set of orthogonal matrices appear. These matrices have certain properties such that if matrices of such properties exist in higher dimensions, then using them one can construct metrologically useful PPT states. The second family of states has been constructed this way.  We can see that for a given $d,$ there are many possibilities to construct such states. Besides $d=3$ we give explicit examples of this construction for $d=2^n$ with $n\ge 1$. We believe that such states exist for other dimensions as well.  For $d\ge3,$ the density matrix and its partial transpose have the same rank and even the same eigenvalues as the states presented in Ref.~\cite{Toth18}. We also show that the algorithm maximizing the quantum Fisher information presented in Ref.~\cite{Toth18} finds such states, and these states are fixed points of the algorithm.

\begin{figure}[t!]
\includegraphics[width=8cm]{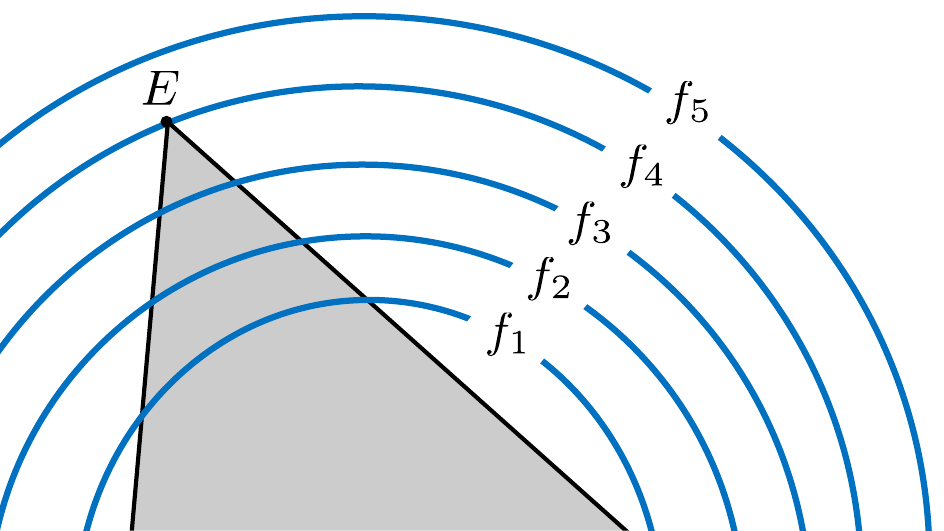}
\caption{PPT states and quantum Fisher information in a section of the state space. (grey area) Points corresponding to PPT states. 
(white area) Points corresponding to non-PPT states. 
(solid curves) Contour plot of the quantum Fisher information ${\cal F}_Q[\varrho,\mathcal H].$ In the figure, $f_n<f_{n+1}$ holds.
(point E) Point corresponding to an extremal state of the PPT set maximizing the quantum Fisher information within the set.  Note that it is also imaginable that the quantum Fisher information is maximized by states that are on the boundary of the set, but not extremal, or by states that are within the set. Still, at least one of the states with a maximal quantum Fisher information must be extremal. 
} \label{fig:pptextremal}
\end{figure}

We will present arguments also for the following statement.

{\bf Conjecture 1.}---For bipartite PPT quantum states of $2d\times 2d$ systems, and for the Hamiltonian given in Eq.~(\ref{eq:Ham}), the quantum Fisher information given in Eq.~(\ref{eq:qfinfo}) is maximal. 

Connected to these, we will investigate the question whether the states we present are extremal within the set of PPT states, as shown in Fig.~\ref{fig:pptextremal}, and we will show the following.

{\bf Observation 2}.---For both families of states, 
\begin{equation}
{\cal F}_Q[\rhopptn,H]=4 (\Delta H)^2_{\rhopptn}
\label{eq:qfinfo2}
\end{equation}
holds and for the expectation value of the Hamiltonian in Eq.~(\ref{eq:Ham})
\begin{equation}
\langle H \rangle_{\rhopptn}=0\label{eq:FQVAR2_2B}
\end{equation}
holds.

The relation in Eq.~(\ref{eq:qfinfo2}) is evidently true for pure states. It is intriguing that it is also true for the mixed states presented in this paper.

Finally, we also prove a statement concerning the metrological performance of the two families of the states mixed with white noise. 

\begin{figure}[t!]
\includegraphics[width=8cm]{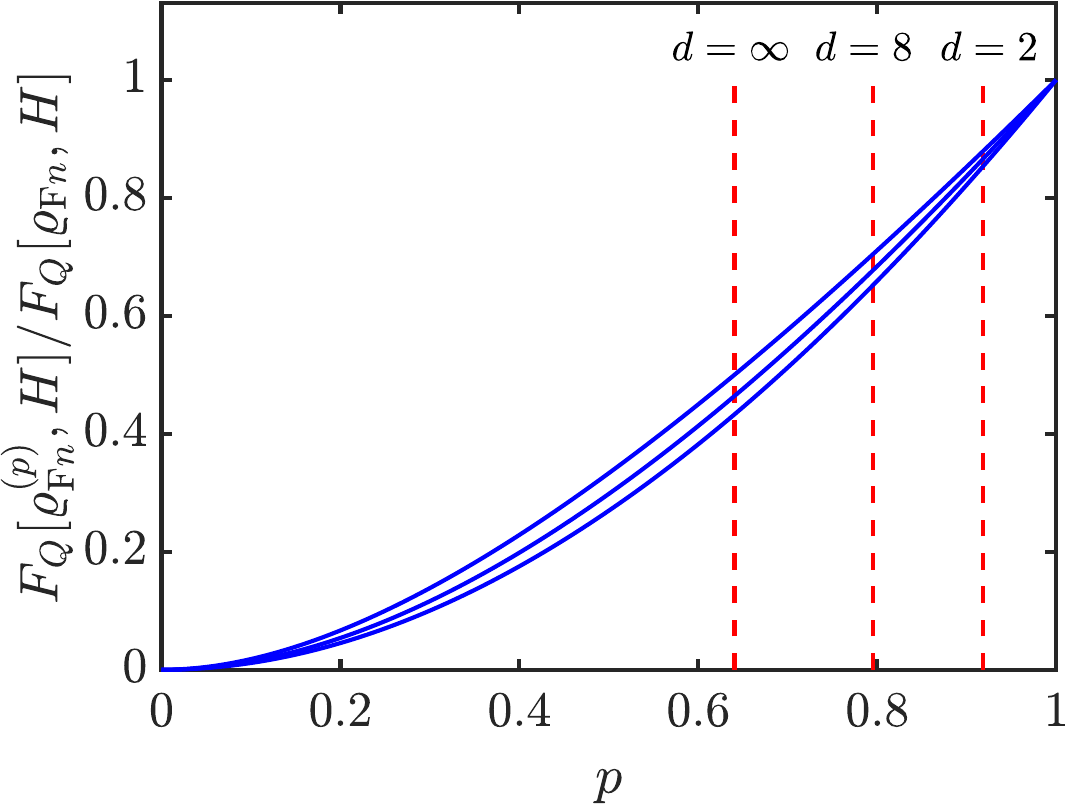}
\caption{(solid) The quantum Fisher information for the $\rhopptn$ state of size $2d\times2d$ mixed with white noise given in Eq.~(\ref{eq:fqwhitenoise12}) as a function of $p$ for (bottom to top) $d=2,8,\infty.$ (dashed) Noise limits for metrological usefulness for the same dimensions. If $p$ is larger than this bound then the state is more useful for metrology than separable states. It can be seen that for large $d,$ the state remains useful even for relatively small $p.$ 
} \label{fig:FQnoisy}
\end{figure}

{\bf Observation 3.}---If we mix the quantum state $\rhopptn$ with white noise,
\begin{equation}
\label{eq:whitenoise12}
\rhopptfnp=p \rhopptn + (1-p) \frac{\openone}{4d^2},
\end{equation}
the quantum Fisher information is given as
\begin{equation}
\label{eq:fqwhitenoise12}
{\cal F}_Q[\rhopptfnp,H]= \frac{2p_1p^2}{(2p_1-1)p+1}{\cal F}_Q[\rhopptn,H],
\end{equation}
where ${\cal F}_Q[\rhopptn,H]$ is given in Eq.~(\ref{eq:qfinfo}). The constant $p_1,$ which will have a central role in this paper, is defined as
\begin{equation}
p_1=\sqrt{d}/(1+\sqrt{d}).\label{eq:p1def} 
\end{equation}
We plotted Eq.~(\ref{eq:fqwhitenoise12}) in Fig.~\ref{fig:FQnoisy}. We also indicated the bounds for $p$ for states that are still useful for metrology. We will present these bounds when discussing robustness of metrological usefulness.

The structure of this paper is as follows. In Sec.~\ref{sec:mot}, we give further motivations for our work. 
In Sec.~\ref{sec:QFIreview}, we review quantum metrology in linear interferometers and also define quantum Fisher information, a key quantity in quantum metrology. 
In Sec.~\ref{sec:QFIppt}, the first family of ($2d\times 2d$)-dimensional PPT states described in Ref.~\cite{Badziag} is presented. We calculate the quantum Fisher information for this construction. 
In Sec.~\ref{sec:TVnum}, we present the second construction based on the numerical results of Ref.~\cite{Toth18} and give the quantum Fisher information for this class of states. 
In Sec.~\ref{sec:GenHam}, we argue that the state we present provides the largest metrological gain possible, even if we consider other Hamiltonians.
In Sec.~\ref{sec:disc}, we discuss the conjecture that the presented states have the possible best metrological performance among bipartite PPT states of a given dimension.
In the Appendices, we present some details of our calculations. 

\section{Motivation: Entanglement and its use in quantum information processing}\label{sec:mot}

Entanglement is an important resource in quantum information science~\cite{horo_review,GT_review}. For instance, all entangled states have been shown to be useful in channel discrimination~\cite{PW09}. On the other hand, there are tasks in which entanglement is required, however, certain entangled states are not useful. 

(i) It is known that entanglement is required to generate Bell nonlocal correlations~\cite{brunner_review,scarani_book}. However, there exist mixed entangled quantum states that admit a local hidden variable model~\cite{werner,barrett}.  Then, such states do not give any advantage over separable states in Bell nonlocal games \cite{Cleve2004}.

(ii) It is also known that entanglement is a necessary ingredient to enhance precision in quantum metrology~\cite{FisherM1,FisherM2}. However, there exist even highly entangled multipartite states that are not better than separable states metrologically~\cite{HGS10}. 

To gain more insight about the role of entangled states in quantum information and particularly in quantum metrology, it is interesting to investigate the metrological usefulness of states that are very weakly entangled. In this paper, we study states that have a positive partial transpose~\cite{PPT}, which are a class of weakly entangled states that play a central role in quantum information science. It is not possible to extract pure singlets from these states even if multiple copies are in our disposal, and we can perform arbitrary local operations and classical communications (LOCC) on these copies~\cite{ED}. Since the entanglement of these states cannot be distilled into singlets, and is trapped in a sense, such states are called bound entangled~\cite{BE}.  These states are not only weakly entangled but are also very mixed, i.e., they contain a large amount of noise. 

Based on these, questions arose about how bound entangled states can be used in quantum information processing. For instance, Peres asked whether these states could violate a Bell inequality~\cite{peres99}. After a long search for examples, his question has also been answered affirmatively: a $3\times 3$ PPT entangled state has been found that violates a specific bipartite Bell inequality with well-chosen measurements~\cite{VB}. More recently, the dimensionality of such Bell nonlocal PPT entangled states has been extended from $3\times 3$ to arbitrary high dimensions~\cite{yuoh,PV17}. 

It also turned out that bipartite PPT entangled states can even have a high Schmidt-rank, which is typically  characteristic of strongly entangled states~\cite{Huber2018} (see also Refs.~\cite{Pal2019,Cariello2019}).

Finally, a recent finding most relevant for this paper is that bound entangled states turn out to be useful in quantum metrology~\cite{czekaj,Toth18}. Metrologically useful bipartite PPT entangled states up to dimension $12\times 12$ have been found numerically in Ref.~\cite{Toth18}. In this paper, we investigate how useful bound entangled states could be for metrology, if we can have large dimensions. 

\section{Metrology and quantum Fisher information}\label{sec:QFIreview}

In this section, we review the basic notions of quantum metrology, and their relations to entanglement theory. For reviews on quantum metrology see Refs.~\cite{Petz08,Paris09,TA14,DJ15,PS18}.

In the most fundamental quantum metrological scenario, a probe state $\varrho$ undergoes a time evolution,  
\begin{equation}
\varrho(\theta)=U_{\theta}\varrho U_{\theta}^\dagger
\end{equation}
 with the unitary dynamics defined as
\begin{equation}
 U_{\theta}=\exp(-{\bf i} \Hop\theta),
 \end{equation}
 where ${\bf i}$ is the imaginary unit, $\varrho(\theta)$ depends on the small angle $\theta$ and the Hamiltonian operator $\Hop$. A crucial question of metrology is, how precisely one can estimate the value of $\theta$ by measuring the state $\varrho(\theta)$. Here we focus on bipartite systems, in which case the Hamiltonian is written as 
 \begin{equation}
 \Hop=\Hop_1+\Hop_2, \label{eq:H12}
  \end{equation}
 where $\Hop_1$ and $\Hop_2$ are single-particle operators. In this setup the precision of the estimation -- assuming any type of measurements -- is limited by the famous Cram\'er-Rao bound as follows \cite{Metrology1,Metrology2,Metrology3}
\begin{equation}
(\Delta\theta)^2\ge \frac{1}{\nu {\cal F}_Q[\varrho,\Hop]},
\end{equation}
where the quantity ${\cal F}_Q$ is the quantum Fisher information given in Eq.~(\ref{eq:FQ}),
and $\nu$ is the number of repeated measurements.

The quantum Fisher information ${\cal F}_Q[\varrho,\Hop]$ is convex in $\varrho,$ and it is bounded from above as \cite{Paris09,TA14,DJ15,PS18}
 \begin{equation}
{\cal F}_Q[\varrho,\Hop]\le 4 (\Delta \Hop)^2_{\varrho}. \label{eq:FQ4VAR_general}
\end{equation}
The inequality in Eq.~(\ref{eq:FQ4VAR_general}) is saturated if $\varrho$ is pure. Interestingly,  the mixed states $\rhopptn$ presented in this paper also saturate Eq.~(\ref{eq:FQ4VAR_general}) for $\mathcal H=H$, as indicated by Observation 2, and will be proven later.

We say that a quantum state is metrologically useful, if it can outperform every separable state in a metrological task defined by a fixed Hamiltonian $\Hop$ \cite{PS09}. That is, 
\begin{equation}
{\cal F}_Q[\varrho,\Hop]>\max_{\varrho_{\rm sep}}{{\cal F}_Q[\varrho_{\rm sep},\Hop]}\equiv {\cal F}_Q^{\rm (sep)}(\Hop).
\end{equation}
For bipartite systems there is an explicit formula for ${\cal F}_Q^{\rm (sep)}(\Hop)$ above given by
\begin{equation}
\sum_{n=1}^2 [\lambda_{\max}(\Hop_n)-\lambda_{\min}(\Hop_n)]^2,
\label{eq:FQsep}
\end{equation}
where $\lambda_{\min}(\mathcal H_n)$ and $\lambda_{\max}(\Hop_n)$ denote the minimum and maximum eigenvalues of $\Hop_n$,  respectively (see e.g., Refs.~\cite{Toth18,Ciampini16} for the derivation of the formula in Eq.~(\ref{eq:FQsep}) and its generalizations for multipartite systems). 

Based on Eq.~(\ref{eq:FQsep}), straightforward calculations show that the  maximum of the quantum Fisher information for separable states is 
\begin{equation}
{\cal F}_Q^{\rm (sep)}(H)=8,
\label{eq:fqsepH}
\end{equation}
where $H$ is given in Eq.~(\ref{eq:Ham}). We will use this bound throughout this paper.

Finally, we define the metrological gain with respect to this Hamiltonian $\Hop$ as 
\cite{Toth2020Activating}
\begin{equation}
g_{\Hop}(\varrho)=\frac{{\cal F}_Q[\varrho,\Hop]}{{\cal F}_Q^{\rm (sep)}(\Hop)}.
\label{eq:metrogain}
\end{equation}
Clearly, we are interested in quantum states for which $g_{\Hop}$ is large. One can even define the metrological gain maximized over local Hamiltonians as
\begin{equation}
g (\varrho)=\max_{\Hop} g_{\Hop}(\varrho)
\end{equation}
For a given system size and for a given set of quantum states, we can ask which $\varrho$ and local $\Hop$ make the largest gain possible.

\section{Quantum Fisher information of PPT states based on private states}\label{sec:QFIppt}

In this section, we calculate the quantum Fisher information of the $(2d\times 2d)$-dimensional PPT states $\rhopptfirst$ and we also give the noise robustness of the metrological advantage of these states.  

\subsection{Spectral decomposition of the state}

Let us first calculate the eigenvalues and eigenstates of $\rhopptfirst$. 

{\bf Definition 1}.---In matrix notation, the states $\rhopptfirst$ 
 can be written as 
\begin{equation}
\rhopptfirst=\frac{1}{2}
\left[{\begin{array}{cccc}
p_1\sqrt{XX^\dag}&0&0&p_1X\\
0&p_2\sqrt{YY^\dag}&p_2Y&0\\
0&p_2Y^\dag&p_2\sqrt{Y^\dag Y}&0\\
p_1X^\dag&0&0&p_1\sqrt{X^\dag X}\\
\end{array} } \right],
\label{eq:rhomat}
\end{equation}
where the two matrices with a unit trace norm acting on $A'B'$ are defined as
 \begin{align}
X=&\frac{1}{d\sqrt{d}}\sum_{i,j=0}^{d-1}u_{ij}|ij\rangle\langle ji|,\nonumber\\
Y=&\sqrt{d}X^\Gamma=\frac{1}{d}\sum_{i,j=0}^{d-1}u_{ij}|ii\rangle\langle jj|.
\label{eq:XY}
\end{align}
Here $\Gamma$ denotes partial transposition in terms of Bob.
Based on Eq.~(\ref{eq:rhomat}), the density matrix of the $\rhopptfirst$ state can also be expressed as
\begin{align}
\rhopptfirst&=\frac{p_1}{2d^2}\sum_{i,j=0}^{d-1}\left(|00ij\rangle\langle 00ij|+|11ij\rangle\langle 11ij|\right)\nonumber\\
&+\frac{p_1}{2d\sqrt{d}}\sum_{i,j=0}^{d-1}\left(u_{ij}|00ij\rangle\langle 11ji|+u^*_{ij}|11ji\rangle\langle 00ij|\right)\nonumber\\
&+\frac{p_2}{2d}\sum_{i=0}^{d-1}\left(|01ii\rangle\langle 01ii|+|10ii\rangle\langle 10ii|\right)\nonumber\\
&+\frac{p_2}{2d}\sum_{i,j=0}^{d-1}\left(u_{ij}|01ii\rangle\langle 10jj|+u^*_{ij}|10jj\rangle\langle 01ii|\right),
\label{eq:rhopri}
\end{align}
where the order of subsystems is $ABA'B',$ 
the $p_1$ probability is given in Eq.~(\ref{eq:p1def}), we define also
\begin{equation}
p_2=1-p_1=1/(1+\sqrt{d}), \label{eq:p2def}
\end{equation}
and the $u_{ij}$ are matrix elements of a unitary operator acting on a $d$-dimensional space fulfilling
 \begin{align}
|u_{ij}|=1/\sqrt{d}\label{eq:uij}
\end{align}
for all $i,j.$  Such an operator exists for all $d$, and the one corresponding to the quantum Fourier transform is appropriate, that is, 
 \begin{align}
u_{ij}=\frac 1 {\sqrt{d}} e^{{\bf i}\frac{2\pi}{d}ij}.
\end{align}
For even $d$ values the $d\times d$ Hadamard matrix multiplied by $1\sqrt{d}$ is also a good choice whenever it exists. It gives a real density matrix. In  Appendix~\ref{sec:privatePPT}, we present the derivation of Eqs.~(\ref{eq:rhomat}) and (\ref{eq:rhopri}) based on the private states of Ref.~\cite{Badziag}. In Appendix~\ref{sec:appendix_bound_private}, we show that the $4\times 4$ bound entangled states of Ref.~\cite{Toth18} are of this form. Note that throughout this paper we start the indices of the rows and columns of matrices from zero, which simplifies many of our formulas.

It is quite straightforward to check that the eigenvectors and the eigenvalues of the density matrix $\rhopptfirst$ above are the following, where we remember that the condition in Eq.~(\ref{eq:uij}) must be fulfilled. There are $d^2$ eigenvectors belonging to the eigenvalue 
\begin{equation}
\Lambda_v=p_1/d^2\label{eq:lambdav}
\end{equation}
 given as
\begin{equation}
|v_{ij}\rangle=\frac{1}{\sqrt{2}}\left(|00ij\rangle+\sqrt{d}u^*_{ij}|11ji\rangle\right).
\label{eq:vkl}
\end{equation}
There are $d$ vectors belonging to the eigenvalue 
\begin{equation}
\Lambda_w=p_2/d\label{eq:lambdaw}
\end{equation}
 defined as
\begin{equation}
|w_{i}\rangle=\frac{1}{\sqrt{2}}\left(|01ii\rangle+\sum_{j=0}^{d-1}u^*_{ij}|10jj\rangle\right).
\label{eq:wk}
\end{equation}
All vectors orthogonal to the ones above belong to the zero eigenvalue. These include vectors having a form like that of $\ket{v_{ij}}$ but with a subtraction sign instead of the addition sign, given as
\begin{equation}
|v_{ij}^-\rangle=\frac{1}{\sqrt{2}}\left(|00ij\rangle-\sqrt{d}u^*_{ij}|11ji\rangle\right).\label{eq:wminus}
\end{equation}
Analogously, from $\ket{w_{i}}$ one can obtain further eigenvectors with a zero eigenvalue
\begin{equation}
|w^-_{i}\rangle=\frac{1}{\sqrt{2}}\left(|01ii\rangle-\sum_{j=0}^{d-1}u^*_{ij}|10jj\rangle\right).
\label{eq:wkminus}
\end{equation}
States $|01ij\rangle$ and $|10ij\rangle$ for all $i\neq j$ also belong to the zero eigenvalue.
This collection of state vectors span the $4d^2$ dimensional Hilbert space of the system.
The eigenvectors with nonzero eigenvalues are $|v_{ij}\rangle$ and  $|w_{i}\rangle.$ Hence,  the rank of the state is given in Eq.~(\ref{eq:rhorank1}). 

Connected to these, we give some details in the Appendices. In Appendix~\ref{sec:alternative}, we review the numerical algorithm maximizing the quantum Fisher information, and show an improved version of the algorithm presented in Ref.~\cite{Toth18}. In Appendix~\ref{sec:mapping}, we show that the states $\rhopptfirst$ are not fixed points of the algorithm given in Ref.~\cite{Toth18} of maximizing the quantum Fisher information. In Appendix~\ref{sec:appendix4by4state}, we show that for the special case of $d=2$ the state $\rhopptfirst$ in Eq.~(\ref{eq:rhopri}) can be written as an equal mixture of symmetric and antisymmetric states.

Finally, one can see by inspection that the states in Eq.~(\ref{eq:rhomat}) are PPT invariant, that is,  
\begin{equation}
\rhopptfirst=(\rhopptfirst)^{\Gamma}, \label{eq:rho_eq_rhopt}
\end{equation}
where $\Gamma$ denotes partial transpose according to $BB'.$ Thus, the state $\rhopptfirst$ has positive partial transpose indeed. We now prove Eq.~(\ref{eq:rho_eq_rhopt}). The partial transposition replaces $p_1X$ by $p_2Y^\Gamma$  in Eq.~(\ref{eq:rhomat}), while it replaces  $p_2Y$ by $p_1X^\Gamma.$ The other elements in the $4\times 4$ block matrix do not change, since 
\begin{align}
\sqrt{XX^\dag}=&\sqrt{X^\dag X}=\frac{1}{d^2}\sum_{i,j=0}^{d-1}|ij\rangle\langle ij|,\nonumber\\
\sqrt{YY^\dag}=&\sqrt{Y^\dag Y}=\frac{1}{d}\sum_{i=0}^{d-1}|ii\rangle\langle ii|
\label{eq:XXYY}
\end{align}
are invariant under a partial transposition with respect to $B'.$ We have to remember that $p_k$ are given by  Eqs.~(\ref{eq:p1def}) and (\ref{eq:p2def}).  We also have to remember the relation between $X$ and $Y$ given in Eq.~(\ref{eq:XY}). Due to these $p_1X=p_2Y^\Gamma$ and $p_2Y=p_1X^\Gamma$ hold, and the  partial transpose does not change $\rhopptfirst.$

\subsection{Quantum Fisher information}

We now calculate the quantum Fisher information for the probe state in Eq.~(\ref{eq:rhopri}) using the formula in Eq.~(\ref{eq:FQ}). 

{\bf Observation 4.}---For the state $\rhopptfirst,$ for the term in the formula of the quantum Fisher information in Eq.(\ref{eq:FQ}), we have
\begin{eqnarray}
&&\langle \mu \vert {H} \vert \nu \rangle =\nonumber\\
&&\quad\quad\quad\quad\begin{cases}
2,& \text{if } \ket{\mu}=\ket{v_{ij}} \text{ and } \; \ket{\nu}=\ket{v^-_{ij}} \\
   &  \text{or } \ket{\nu}=\ket{v_{ij}} \text{ and } \; \ket{\mu}=\ket{v^-_{ij}}, \\
0,& {\rm otherwise},
\end{cases}\label{eq:Hmunu_rho2}
\end{eqnarray}
where $0\le i,j\le d-1.$
Here $\vert \mu \rangle$ and $\vert \nu \rangle$ denote the eigenvectors of $\rhopptfirst$ listed before.
They include all $\ket{v_{ij}}$'s and all $\ket{v_{ij}^-}$'s.
The Hamilton operator used is 
given in Eq.~(\ref{eq:Ham}), which we repeat here for clarity
\begin{equation}
H=\sigma^z_A\otimes \openone_B\otimes  \openone_{A'B'}+ \openone_A\otimes\sigma^z_B\otimes  \openone_{A'B'}.
\end{equation}
This is the same Hamiltonian operator that appears in Ref.~\cite{Toth18} for two-qudit states, apart from some trivial rearrangement of the qudits.  

{\it Proof.}---From Eq.~(\ref{eq:Ham}) follows that 
\begin{subequations}
\begin{eqnarray}
H|00ij\rangle&=&+2|00ij\rangle, \\
H|01ij\rangle&=&0, \\
H|10ij\rangle&=&0, \\
H|11ij\rangle&=&-2|11ij\rangle,
\end{eqnarray}\label{eq:eig}\end{subequations}
for all $i,j.$ 
The vectors appearing in Eq.~(\ref{eq:eig}) are clearly eigenvectors of $H$.  From Eqs.~(\ref{eq:vkl}), (\ref{eq:wminus}) and (\ref{eq:eig}), it follows that
\begin{equation}
H|v_{ij}\rangle=2|v^-_{ij}\rangle.
\label{eq:Avkl}
\end{equation}
Equation~(\ref{eq:Avkl}) proves the first two lines of Eq. (36). It also follows that $|v_{ij}\rangle$ and $|v_{ij}^-\rangle$ give nonzero contribution to the Fisher information only with each other, that is if $|\mu\rangle=|v_{ij}\rangle$ and $|\mu\rangle\ne|v_{ij}^-\rangle$ then  $\langle\mu|H|\nu\rangle=0.$ Moreover,  if $|\mu\rangle=|v_{ij}^-\rangle$ and $|\mu\rangle\ne|v_{ij}\rangle$ then  also $\langle\mu|H|\nu\rangle=0.$ Note that the vectors $|v_{ij}\rangle$ and  $|v_{ij}^-\rangle$  represent $2d^2$ eigenvectors of $\rhopptfirst$, and they live in the space spanned by  $\ket{00ij}$ and $\ket{11ij}.$

There are other $2d^2$ eigenvectors  of $\rhopptfirst$ in the space of $\ket{01ij}$ and $\ket{10ij},$ which include $|w_{i}\rangle$ and $|w_{i}^-\rangle.$  The states $\ket{01ij}$ and $\ket{10ij}$ are eigenvectors of $H$ with eigenvalue zero according to Eq.~(\ref{eq:eig}), therefore $\langle\mu|H|\nu\rangle=0$, whenever either $|\mu\rangle$ or $|\nu\rangle$ is from this subspace. Since we covered all $4d^2$ eigenvectors of the density matrix, this concludes the proof of Observation 4. $\qed$

Note that  Eq.~(\ref{eq:qfinfo}) approaches 16 for large $d$, which is the theoretical maximum one can achieve with the Hamiltonian~(\ref{eq:Ham}), see Appendix~\ref{sec:appendix0}. 

Next, we prove an interesting statement concerning the quantum Fisher information and the variance of $H$ computed for the quantum states $\rhopptfirst.$

{\it Proof of Observation 2 for $\rhopptfirst$.}---From Observation 4, it follows that
for the eigenstates of $\rhopptfirst$ the relation
\begin{equation}
\lambda_\mu \lambda_\nu  \langle \mu \vert H \vert \nu\rangle =0\label{eq:lambdakl}
\end{equation}
holds, where $\ket \mu$ and $\ket \nu$ are eigenstates of  $\rhopptfirst$ and $\lambda_\mu$ and $\lambda_\nu$ are corresponding eigenvalues. This can be seen noting that Eq.~(\ref{eq:lambdakl}) expresses the fact that if $\ket \mu$ and $\ket \nu$  are eigenvectors with a nonzero eigenvalue, then
\begin{equation}
 \langle \mu \vert H \vert \nu \rangle =0.\label{eq:lambdakl2}
\end{equation}
From Eq.~(\ref{eq:lambdakl2}), it follows that the expectation value of $H$ is zero as stated in Eq.~(\ref{eq:FQVAR2_2B}).
The formula for the quantum Fisher information in Eq.~(\ref{eq:FQ}) can be rewritten as 
\begin{equation}
\label{eq:FQ2}
{\cal F}_Q[\varrho,{H}]=4\langle H^2\rangle_\varrho - 8\sum_{\mu,\nu}\frac{\lambda_{\mu}\lambda_{\nu}}{\lambda_{\mu}+\lambda_{\nu}} \left|\langle \mu \vert {H} \vert \nu \rangle \right|^{2}.
\end{equation}
From Eqs.~(\ref{eq:lambdakl}) and (\ref{eq:FQ2}), the main statement in Eq.~(\ref{eq:qfinfo2}) follows. $\qed$

As we noted already in Sec.~\ref{sec:QFIreview}, the inequality with the quantum Fisher information and the variance given in Eq.~(\ref{eq:FQ4VAR_general}) is saturated typically by pure states, however, it is also saturated by
$\rhopptfirst.$

Let us see now another consequence of Observation 4. We will show that the quantum Fisher information for a noisy state can easily be given analytically with a simple formula, due to the special form of the Hamiltonian.

{\it Proof of Observation 3 for $\rhopptfirst.$}---Due to Eq.~(\ref{eq:Hmunu_rho2}), it is straightforward to calculate the effect of white noise. The eigenvectors of $\rhopptfirstp$ are obviously the same as that of $\rhopptfirst$, while its eigenvalues will be those of $\rhopptfirst$ multiplied by $p$ and $(1-p)/4d^2$ added to them. $\qed$

We can relate this formula to the quantum Fisher information of a pure state mixed with white noise given as \cite{FisherM1,FisherM2}
\begin{equation}
{\cal F}_Q[\varrho_{p},A]=\frac{p^2}{p+2(1-p)d^{-1}}{\cal F}_Q[\ket{\Psi},A],\label{eq:FQ_iso1}
\end{equation}
where the noisy pure state is given as 
\begin{equation}
\label{eq:whitenoise1}
\varrho_p=p |\Psi\rangle\langle\Psi| + (1-p) \frac{\openone}{d_\Psi^2},
\end{equation}
and $A$ is some operator, $d_\Psi$ is the dimension of the pure state. With $d_\Psi=4 p_1$ Eq.~(\ref{eq:FQ_iso1}) gives the same dependence as Eq.~(\ref{eq:fqwhitenoise12}). For large $d,$ $p_1$ defined in Eq.~(\ref{eq:p1def}) approaches $1$ and  hence $4p_1$ approaches $4$ for large $d.$ Thus, we get the same dependence on white noise as for a pure state with dimension $4.$

\subsection {Metrological gain}
\label{sec:metgain}

The maximum value attainable with separable states using the same Hamiltonian in Eq.~(\ref{eq:Ham}) is 
given in Eq.~(\ref{eq:fqsepH}). Hence, the metrological gain $g_H(\rhopptfirst)$ defined by Eq.~(\ref{eq:metrogain}) is 
\begin{equation}
g_H(\rhopptfirst)=2\frac{\sqrt{d}}{1+\sqrt{d}},
\end{equation}
which, for large $d,$ converges to the theoretical limit of $2$ attainable by two-qubit maximally entangled states.  

\subsection{Robustness against noise}
\label{sec:rob_noise}

In an experiment, the quantum state is never prepared with a perfect fidelity. Thus, it is important  to examine how useful the state remains metrologically, if it is mixed with noise. The resistance to noise can be characterized by the robustness of metrological usefulness, which is just the maximal fraction $r$ of white noise that can be added to the state such that the corresponding quantum Fisher information is still not smaller than the maximum achievable by separable states \cite{Toth18}.

Let us now calculate the robustness of metrological usefulness $r$ of the state $\rhopptfirst.$ For that, we have to look for the $r$ for which
\begin{equation} 
{\cal F}_Q[\rhopptfirstr,H]={\cal F}_Q^{({\rm sep})}\label{eq:robust_eq}
\end{equation}
holds, where the left-hand side of Eq.~(\ref{eq:robust_eq}) is given in Eq.~(\ref{eq:fqwhitenoise12}). Using Eq.~(\ref{eq:FQsep}), the right-hand side of Eq.~(\ref{eq:robust_eq})  is given by Eq.(\ref{eq:fqsepH}). Then, the robustness of metrological usefulness is
\begin{equation}
r=1-\frac{2p_1-1+\sqrt{(2p_1-1)^2+16p_1^2}}{8p_1^2}.
\label{eq:qcrit}
\end{equation}
For large $d,$ the value $p_1$ given in Eq.~(\ref{eq:p1def}) approaches 1, so the robustness of metrological usefulness $r$ approaches 
\begin{equation}
1-(1+\sqrt{17})/8\approx 0.3596.
\end{equation}
In Fig.~\ref{fig:FQnoisy}, we indicated by dashed lines the value of $1-r$ for $d=2,8,$ and $\infty.$

\section{Second family of states and its quantum Fisher information}\label{sec:TVnum}

In this section, we present another class of PPT entangled states, for which Eq.~(\ref{eq:qfinfo}) holds, i.e., their metrological performance depends on their dimension in the same way as in the case of the states discussed in the previous section. The family of states was obtained after we studied the states found numerically  in Ref.~\cite{Toth18}.

{\bf Definition 2.}---The family of states  in terms of their eigenvectors  can be written as
\begin{align}
\rhopptsecond&=\frac{p_1}{d^2}\sum_{i,j=0}^{d-1}|z_{ij}\rangle\langle z_{ij}|+\frac{p_2}{2d}\sum_{i=0}^{d-1}|s_i \rangle\langle s_i|\nonumber\\
&+\frac{p_2}{2d}\sum_{i=0}^{d-1}|10ii\rangle\langle 10ii|,
\label{eq:rho}
\end{align}
where the subscript 2 refers to the  second family of PPT quantum states we consider in this paper. The rank can be read from Eq.~(\ref{eq:rho}) as given in Eq.~(\ref{eq:rhorank2}). The probabilities $p_1$ and $p_2$ are given in  Eqs.~(\ref{eq:p1def}) and (\ref{eq:p2def}), and
\begin{equation}
|z_{ij}\rangle=\frac{1}{\sqrt{2}}\left(|00ij\rangle+\sum_{k=0}^{d-1}Q_{ik}^j|11jk\rangle\right)
\label{eq:zvec}
\end{equation}
for $0\le i,j\le d-1,$ where $Q_{ik}^j$ are orthogonal matrices for all values of $j,$ that is,
\begin{equation}
\sum_{i} Q_{ik}^j Q_{ik'}^j = \delta_{kk'}
\end{equation}
holds for all $j.$ Moreover, $Q_{ik}^j$ also have further properties that will be detailed later. The states $|s_i\rangle$ are orthonormal vectors in the subspace 
\begin{equation}
|01\rangle_{AB}\otimes
\mathcal{H}_{A'}\otimes\mathcal{H}_{B'}, \label{eq:svec}
\end{equation}
which will also be specified later in terms of $Q_{ik}^j$. We will show that with an appropriate choice of the $Q_{ik}^j$ the partial transpose of $\varrho$ is positive semidefinite. 

The partial transpose of the part of $\varrho$ belonging to eigenvalue $p_1/d^2$, as it can be derived from Eq.~(\ref{eq:zvec}) is
\begin{align}
&\left(\sum_{i,j=0}^{d-1}|z_{ij}\rangle\langle z_{ij}|\right)^{\Gamma}=\frac{1}{2}\sum_{i,j=0}^{d-1}\Big(|00ij\rangle\langle\ 00ij|\nonumber\\
&+|11ji\rangle\langle 11ji|+|01ij\rangle\langle q_{ij}|+|q_{ij}\rangle\langle\ 01ij|\Big),
\label{eq:zptr}
\end{align}
where we have introduced the notation 
\begin{equation}
|q_{ij}\rangle\equiv\sum_{k=0}^{d-1} Q_{ij}^k|10kk\rangle,
\label{eq:qvec}
\end{equation}
where $0\le i,j\le d-1.$

We now show what kind of properties $Q_{ij}^k$ need to have. As $|q_{ij}\rangle$ are in the $d$-dimensional subspace spanned by vectors $|10kk\rangle,$  there can not be more than $d$ linearly independent vectors among them. Let matrices $Q_{ij}^k$ be such that there are exactly $d$ such vectors, which are orthogonal to each other, but not necessarily normalized. Moreover, let each $|q_{ij}\rangle$ be either the zero vector, or equal to one of the orthogonal basis vectors with a $+1$ or a $-1$ factor. 

Following these lines, we formulate various relations for the $|q_{ij}\rangle$ and $Q_{ij}^k$ in more detail, which are necessary to construct our bound entangled states.
Let us normalize the orthogonal vectors and let us denote them after normalization by 
 \begin{equation}
|\bar q_{m}\rangle \text{ for } m=0,\dots,d-1. \label{eq:qm}
 \end{equation}
It follows from the properties of $|q_{ij}\rangle $ we required above that the nonzero ones of them are equal to $+1$ or $-1$ times one of the $|\bar q_{m}\rangle,$ up to a constant factor. That is, we require that for every $i,j,$ either $|q_{ij}\rangle$ is zero vector, or there is a $\mu_{ij}$ such that
\begin{equation}
|q_{ij}\rangle=S_{ij}|\bar q_{\mu_{ij}}\rangle \label{eq:barq} 
\end{equation}
is fulfilled, where $S_{ij}\ne 0$. In that case, we also assume that 
 \begin{equation}
 S_{ij} \in \{+D_{\mu_{ij}},-D_{\mu_{ij}}\},\label{eq:Sij}
 \end{equation}
 where $D_{\mu_{ij}}$ is a positive constant we determine later. Let us stress the role of $\mu_{ij}$. It tells us for each $i$ and $j$ for which $|\bar q_{m}\rangle$  Eq.~(\ref{eq:barq}) must hold. Let us now introduce the notation  $\Xi_m$ for the subset of index pairs $(i,j)$ for which $\mu_{ij}=m$ with $S_ {ij}\neq 0,$ that is, 
 \begin{equation}
\Xi_{m}=\{(i,j): |q_{ij}\rangle=S_{ij}|\bar q_{m}\rangle \text{ and } S_ {ij}= \pm D_{m}\}.
\end{equation}
Then, we determine the appropriate value of $D_{m}$ as
 \begin{equation}
D_{m}=\sqrt{d/N_{m}}, \label{eq:S0}
 \end{equation}
 where $N_{m}$ is the number of elements of $\Xi_{m}$.
With this choice
\begin{equation}
\sum_{ij=0}^{d-1}\langle q_{ij}|q_{ij}\rangle=\sum_{m=0}^{d-1}\sum_{(i,j)\in\Xi_{m}}\frac{d}{N_{m}}\langle\bar q_m|\bar q_m\rangle=d^2, 
\end{equation}
which is necessary for consistency, because this is the value one gets using Eq.~(\ref{eq:qvec}) and the fact that $Q_{ij}^k$ is an orthogonal matrix for each fixed $k$.

Using the properties of $Q_{ij}^k$ and the notations presented above we get
\begin{align}
\sum_{i,j=0}^{d-1}|01ij\rangle\langle q_{ij}|=&\sum_{m=0}^{d-1}\sum_{(i,j)\in\Xi_{m}}S_{ij}|01ij\rangle\langle\bar q_m|\nonumber\\
=&\sqrt{d}\sum_{m=0}^{d-1}|t_m\rangle\langle\bar q_m|,
\label{eq:01ijqij}
\end{align}
where the $|t_m\rangle$ states are defined as
\begin{equation}
|t_m\rangle\equiv\frac{1}{\sqrt{d}}\sum_{(i,j)\in\Xi_{m}}S_{ij}|01ij\rangle.
\label{eq:vectm}
\end{equation}
It can be verified that vectors $|t_m\rangle$ are normalized and it is trivial that they are orthogonal to each other. 

Let us now assume that there exist vectors $|s_m\rangle$ 
such that
\begin{equation}
\left(\sum_{i=0}^{d-1}|s_i\rangle\langle s_i|\right)^{\Gamma}=\sum_{m=0}^{d-1}|t_i\rangle\langle t_i|,
\label{eq:st}
\end{equation}
which is a quite non-trivial further requirement imposed on $Q_{ij}^k$. These are the vectors $|s_i\rangle$ appearing in Eq.~(\ref{eq:rho}).

Let us determine now explicitly some of the eigenvectors with a zero eigenvalue. One of them is
\begin{equation}
|z_{ij}^-\rangle=\frac{1}{\sqrt{2}}\left(|00ij\rangle-\sum_{k=0}^{d-1}Q_{ik}^j|11jk\rangle\right),
\label{eq:zvecminus}
\end{equation}
which is like the eigenvector $|z_{ij}\rangle$ given in Eq.~(\ref{eq:zvecminus}), the only difference being the negative sign rather than a positive sign. The relation was similar between $|v_{ij}\rangle$ and $|v_{ij}^-\rangle$ in the case of 
$\rhopptfirst.$  With these, we finished defining $\rhopptsecond.$

Let us now calculate the partial transposition of $\rhopptsecond,$ and show that it is positive semidefinite. Unlike the state $\rhopptfirst,$ the state $\rhopptsecond,$ is not invariant under the partial transposition. Using Eqs.~(\ref{eq:rho}), (\ref{eq:zptr}), (\ref{eq:01ijqij}) and (\ref{eq:st}), the partial transpose of the density matrix can be written as
\begin{align}
(\rhopptsecond)^{\Gamma}&=\frac{p_1}{2d^2}\sum_{i,j=0}^{d-1}(|00ij\rangle\langle 00ij|+|11ji\rangle\langle 11ji|)\nonumber\\
&+\frac{p_1}{2d^{3/2}}\sum_{i=0}^{d-1}(|t_i \rangle\langle\bar q_i|+|\bar q_i\rangle\langle t_i|)\nonumber\\
&+\frac{p_2}{2d}\sum_{i=1}^{d-1}(|t_i\rangle\langle t_i|+|10ii\rangle\langle 10ii|).
\label{eq:rhoptr}
\end{align}
Considering that $p_1/\sqrt{d}=p_2$, and that 
\begin{align}
\sum_i|10ii\rangle\langle 10ii|=\sum_i|\bar q_i\rangle\langle\bar q_i|, 
\end{align}
which follows from the fact that $|10ii\rangle$ and $|\bar q_i\rangle$  for $0\leq i\leq d-1$ are orthonormal bases of the same subspace, we arrive at
\begin{align}
(\rhopptsecond)^{\Gamma}&=\frac{p_1}{2d^2}\sum_{i,j=0}^{d-1}(|00ij\rangle\langle 00ij|+|11ji\rangle\langle 11ji|)\nonumber\\
&+\frac{p_2}{d}\sum_{i=0}^{d-1}\left[\frac{1}{\sqrt{2}}(|t_i\rangle+|\bar q_i\rangle)(\langle t_i|+\langle\bar q_i|)\frac{1}{\sqrt{2}}\right].
\label{eq:rhoptr2}
\end{align}

Let us calculate now the eigenvalues and the rank of $(\rhopptsecond)^{\Gamma}.$ We now show that Eq.~(\ref{eq:rhoptr2}) is an eigendecomposition of the partial transposition.  In the first sum, we can see the $| 00ij\rangle$ and $| 11ji\rangle$ vectors, which are all pairwise orthogonal to each other and they have a unit norm. The $|t_i\rangle$ vectors are defined in Eq.~(\ref{eq:vectm}). They are pairwise orthogonal to each other and they have a unit norm, which can be seen based on Eqs.~(\ref{eq:Sij}) and (\ref{eq:S0}). Since they are in the $|01ij\rangle$ subspace, they are also orthogonal to all previous vectors. The $|\bar q_i\rangle$ vectors given in  Eq.~(\ref{eq:barq}) are pairwise orthogonal to each other. They have a unit norm. Since they are in the $|10ij\rangle$ subspace, they are also orthogonal to all previous vectors. The vectors 
\begin{align}
(|t_i\rangle+|\bar q_i\rangle)/\sqrt2
\end{align}
also have a unit norm, they are pairwise orthogonal, and they are orthogonal to all previous basis vectors. All vectors in the decomposition in Eq.~(\ref{eq:rhoptr2}) have a unit norm, hence the eigenvalues can be clearly seen. The eigenvalue $p_1/(2d^2)$ is $2d^2$-times degenerate, and  the eigenvalue $p_2/d$ is $d$-times degenerate. Based on these, the rank can be read from Eq.~(\ref{eq:rhoptr2}) as given in Eq.~(\ref{eq:rhorank2pt}). These prove that Eq.~(\ref{eq:rhoptr2}) is an eigendecomposition. Hence,   $(\rhopptsecond)^{\Gamma}$ is positive semidefinite, indeed.

For this family of states we have constructed analytical examples only for $d=3$ and $2^n.$ However, we believe that such states exist for all $d$. The numerical procedure looking for states with the maximal quantum Fisher information described in  Ref.~\cite{Toth18}  seems to find such states. (See Appendices~\ref{sec:alternative} and ~\ref{sec:mapping}.) 

We found the analytical $d=3$ state given in Appendix~\ref{sec:d3case} by analyzing the corresponding state obtained numerically in Ref.~\cite{Toth18}. We have recognized the important features of this solution, which made it possible to construct the present family of states for larger dimensions. The construction is based on a set of orthogonal matrices $Q_{ij}^k$ having certain non-trivial properties. The matrices appearing in the $d=3$ solution are block diagonal with $2\times 2$ and $1\times 1$ blocks. The $2\times 2$ blocks in the three orthogonal matrices are characterized by angles corresponding to a regular triangle. 

For higher dimensions we have found explicit analytical examples of this family for $d=2^n$ for any $n>0$ which we give in the Appendix~\ref{sec:devcase}. In these cases, $Q_{ij}^k$ are tensor products of 2-dimensional unit matrices and Pauli $X$ matrices in every order. 

Next, we calculate the quantum Fisher information for our state. For that, first we need to obtain the matrix elements of the Hamiltonian in the eigenbasis of the density matrix.

{\bf Observation 5.}---For the state $\rhopptsecond,$ for the term in the formula of the quantum Fisher information in Eq.~(\ref{eq:FQ}), we have
\begin{eqnarray}
&&\langle \mu \vert {H} \vert \nu \rangle =\nonumber\\
&&\quad\quad\begin{cases}
2,& \text{if } \ket{\mu}=\ket{z_{ij}} \text{ and } \; \ket{\nu}=\ket{z^-_{ij}} \\
   &  \text{or } \ket{\nu}=\ket{z_{ij}} \text{ and } \; \ket{\mu}=\ket{z^-_{ij}}, \\
0,& {\rm otherwise},
\end{cases}
\end{eqnarray}
where $0\le i,j\le d-1.$ Here $\vert \mu \rangle$ and $\vert \nu \rangle$ denote the eigenvectors of $\rhopptsecond.$ We assume that they include all eigenvectors we listed before. They include all $\ket{z_{ij}}$'s and all $\ket{z_{ij}^-}$'s.

{\it Proof.}---The proof is basically the same as the proof of Observation 4.  The eigenvectors $|z_{ij}\rangle$ play the same role now as $|v_{ij}\rangle$ [see Eqs.~(\ref{eq:rhopri}) and (\ref{eq:vkl})] shown before.  From Eqs.~(\ref{eq:zvec}), (\ref{eq:zvecminus}) and (\ref{eq:eig}), it follows that
\begin{equation}
H|z_{ij}\rangle=2|z^-_{ij}\rangle.
\label{eq:Avkl2}
\end{equation}
Like in the case of Observation 4, this proves the first two lines of Observation 5, and that $|z_{ij}\rangle$ and $|z_{ij}^-\rangle$ give nonzero contribution to the Fisher information only with each other. All other eigenstates of $\rho_{F2}$, either with eigenvalue $p_2/2d$ or zero, live in the subspace belonging to the zero eigenvalue of $H$ [see Eqs. (\ref{eq:rho}) and (\ref{eq:eig})], therefore, they do not contribute, which concludes the proof of Observation 5. $\qed$

Now we calculate the metrological usefulness of the states presented.

{\it Proof of Observation 1 for $\rhopptsecond.$}---The rank of $\rhopptsecond$ is given in Eq.~(\ref{eq:rhorank2}). It is different from the rank of $\rhopptfirst$ given in Eq.~(\ref{eq:rhorank1}). Nevertheless, the quantum Fisher information with the Hamiltonian in Eq.~(\ref{eq:Ham})  is obviously the same. The calculation can be done analogously to calculations done for $\rhopptfirst.$ Based on Observation 5, in the formula of the quantum Fisher information in Eq.(\ref{eq:FQ}), $2d^2$ of the terms  
\begin{equation}
\left|\langle \mu \vert {H} \vert \nu \rangle \right|^2\label{eq:coeffB}
\end{equation}
are nonzero and they all have the value $4.$ Moreover, in this case, the coefficient of these terms are all
\begin{equation}
\frac{(\lambda_{\mu}-\lambda_{\nu})^{2}}{\lambda_{\mu}+\lambda_{\nu}} = p_1/d^2,
\end{equation}
where $\Lambda_v$ is given in Eq.~(\ref{eq:lambdav}). Hence, we obtain
\begin{equation}
{\cal F}_Q[\rhopptfirst,H]=16 p_1,
\end{equation}
which leads to Eq.~(\ref{eq:qfinfo}). This derivation was almost the same as the derivation of Observation 1 for $\rhopptfirst.$ $\qed$

Next, we prove an interesting statement concerning the quantum Fisher information and the variance of $H$ computed for the quantum states $\rhopptsecond.$

{\it Proof of Observation 2 for $\rhopptsecond$.}---From Observation 5, it follows that
for the eigenstates of $\rhopptsecond$ the relation
\begin{equation}
\lambda_\mu \lambda_\nu  \langle \mu \vert H \vert \nu\rangle =0\label{eq:lambdaklb}
\end{equation}
holds, where $\ket \mu$ and $\ket \nu$ are eigenstates of  $\rhopptsecond$
and $\lambda_\mu$ and $\lambda_\nu$ are corresponding eigenvalues. 
This can be seen noting that the Eqs.~(\ref{eq:lambdaklb}) express the fact that if $\ket \mu$ and $\ket \nu$  are eigenvectors with a nonzero eigenvalue, then
\begin{equation}
 \langle \mu \vert H \vert \nu \rangle =0.\label{eq:lambdakl2b}
\end{equation}
From Eq.~(\ref{eq:lambdakl2b}), it follows that the expectation value of $H$ is zero as stated in Eq.~(\ref{eq:FQVAR2_2B}). The formula for the quantum Fisher information in Eq.~(\ref{eq:FQ}) can be rewritten as in Eq.~(\ref{eq:FQ2}). From Eqs.~(\ref{eq:lambdaklb}) and (\ref{eq:FQ2}), the main statement in Eq.~(\ref{eq:qfinfo2}) follows. $\qed$

As we noted already in Sec.~\ref{sec:QFIreview},  the inequality with the quantum Fisher information and the variance given in Eq.~(\ref{eq:FQ4VAR_general}) is saturated typically by pure states, however, we see that it is also saturated by $\rhopptsecond.$

{\it Proof of Observation 3 for $\rhopptsecond.$}---The proof is analogous to the proof of Observation 3 for $\rhopptfirst.$ $\qed$

Similarly, the metrological gain and the robustness of metrological usefulness is the same for $\rhopptsecond$ as it was  for $\rhopptfirst$  discussed in Sec.~\ref{sec:metgain} and \ref{sec:rob_noise}, respectively.

\section{General Hamiltonian}\label{sec:GenHam}

In this section, we discuss why we considered the Hamiltonian given in Eq.~(\ref{eq:Ham}).  We argue that this Hamiltonian makes possible the largest metrological gain achievable by PPT states. The maximal gain is achieved by the bound entangled states presented in this paper for the Hamiltonian given in Eq.~(\ref{eq:Ham}).

The most general local Hamiltonian is given in Eq.~(\ref{eq:H12}). We will now ask the question, which one makes it possible to have the largest metrological gain among PPT states. Without loss of generality, we can restrict our attention to Hamiltonians of the type
\begin{equation}
H_{\rm gen}=D_{AA'}\otimes  \openone_{BB'}+ \openone_{AA'}\otimes  D_{B'B'}, 
\label{eq:Ham2}
\end{equation}
where the local Hamiltonians are diagonal matrices, since any local Hamiltonian can be transformed into this form by local unitaries, and the maximal metrological gain among PPT states achievable by these Hamiltonians do not change under such transformations. Then, we are also allowed to add $\openone$ times a constant to $D_{AA'},$ and it will not change the metrological properties of the Hamiltonian either. In this way, we can achieve that 
\begin{equation}
\lambda_{\max} (D_{AA'})=-\lambda_{\min} (D_{AA'})=c_1.\label{eq:cond1}
\end{equation}
Similarly, with trivial operation on $D_{B'B'},$ we can achieve that 
\begin{equation}
\lambda_{\max} (D_{BB'})=-\lambda_{\min} (D_{BB'})=c_2.\label{eq:cond2}
\end{equation}

It has been shown in Ref.~\cite{Toth2020Activating} that if we look among the Hamiltonians of the form in Eq.~\eqref{eq:Ham2} with the conditions 
in Eqs.~\eqref{eq:cond1} and \eqref{eq:cond2} for the one with the maximal metrological gain among PPT states then the optimal Hamiltonian will be such that $D_{AA'}$ will have diagonal elements  $\pm c_1,$ while $D_{BB'}$ will have diagonal elements  $\pm c_2.$ Hence, we arrive at a simpler Hamiltonian
\begin{equation}
H_{\rm gen}'=c_1 \tilde D_{AA'}\otimes  \openone_{BB'}+ c_2 \openone_{AA'}\otimes  \tilde D_{B'B'}, 
\label{eq:Ham22}
\end{equation}
where $\tilde D_{AA'}$ and $\tilde D_{B'B'}$ are diagonal matrices with $\pm1$ in their diagonal. When looking for a Hamiltonian with a maximal metrological gain, it is sufficient to consider Hamiltonians of the form in Eq.~(\ref{eq:Ham22}). We can even consider $c_1=1$ without a loss of generality.

\begin{figure}[t!]
\includegraphics[width=4.1cm]{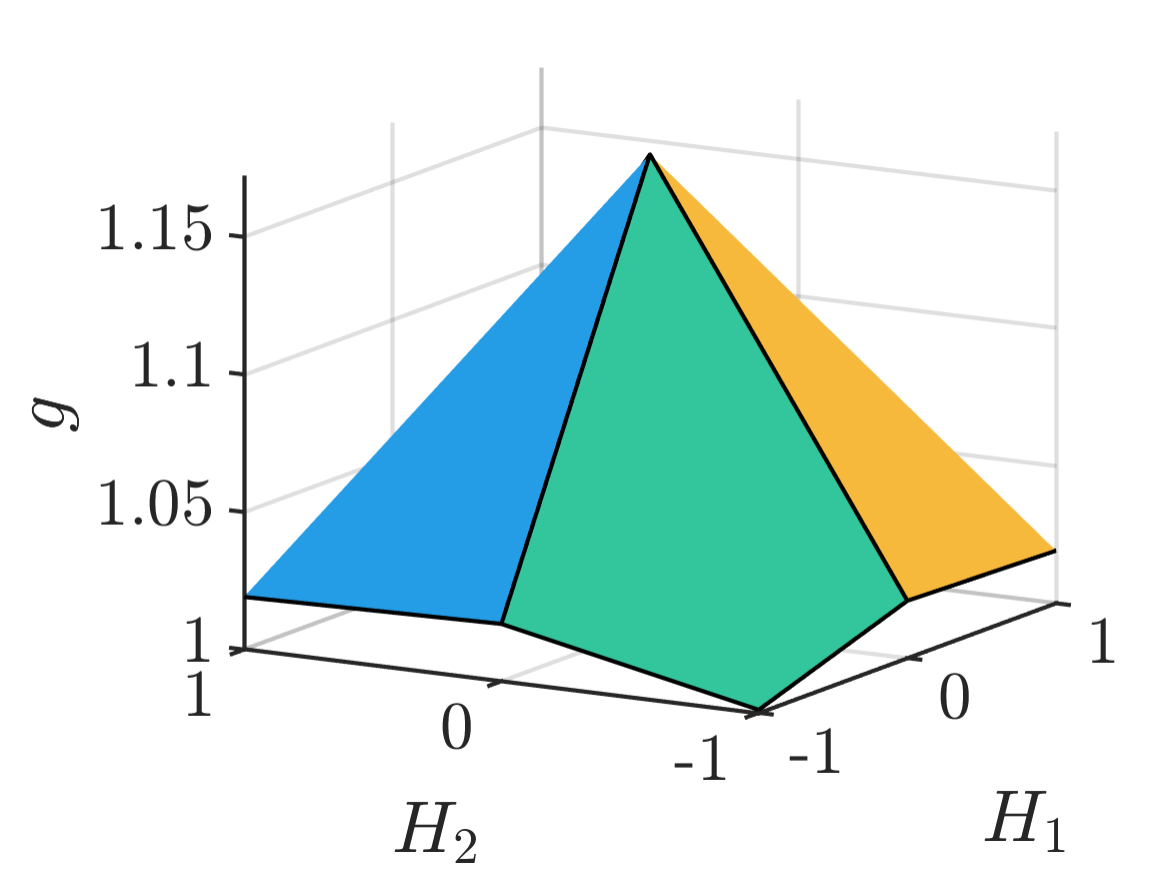}
\includegraphics[width=4.1cm]{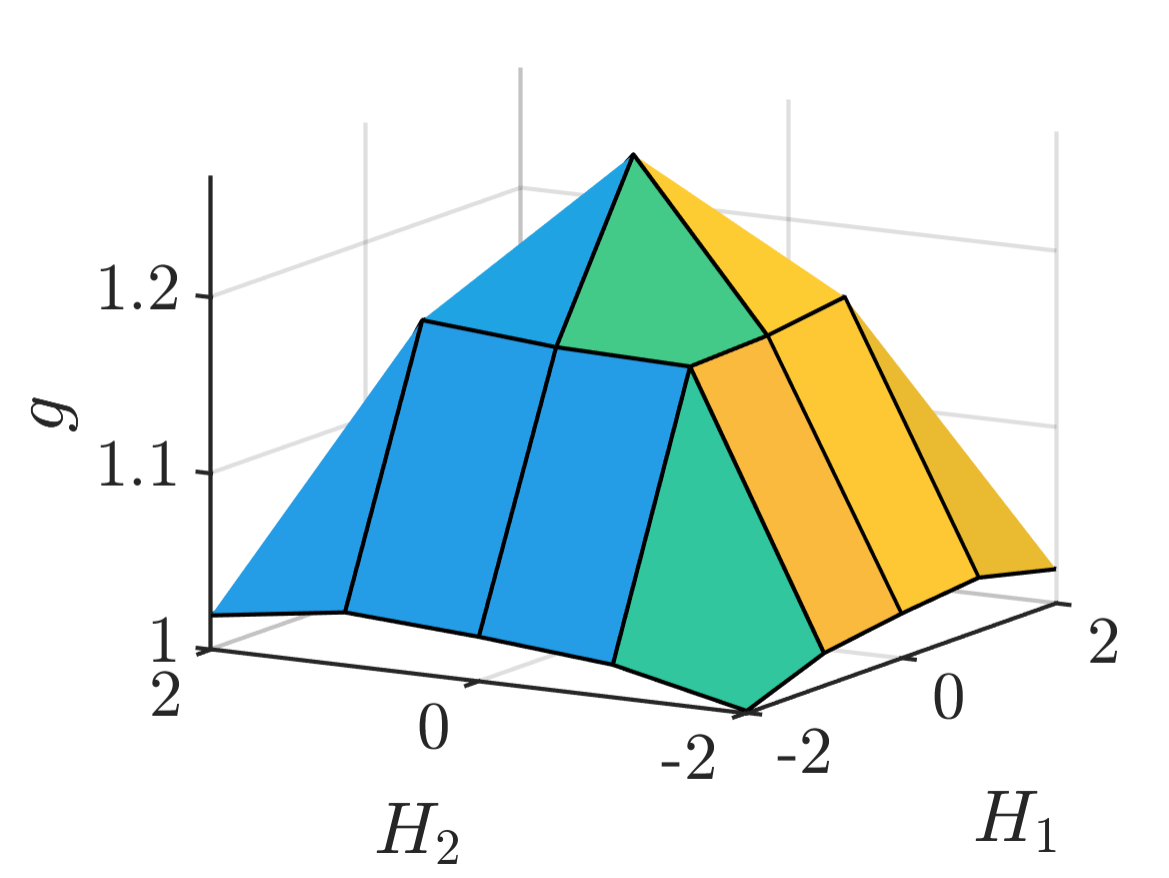}
(a) \hskip3.5cm (b) 

\includegraphics[width=4.1cm]{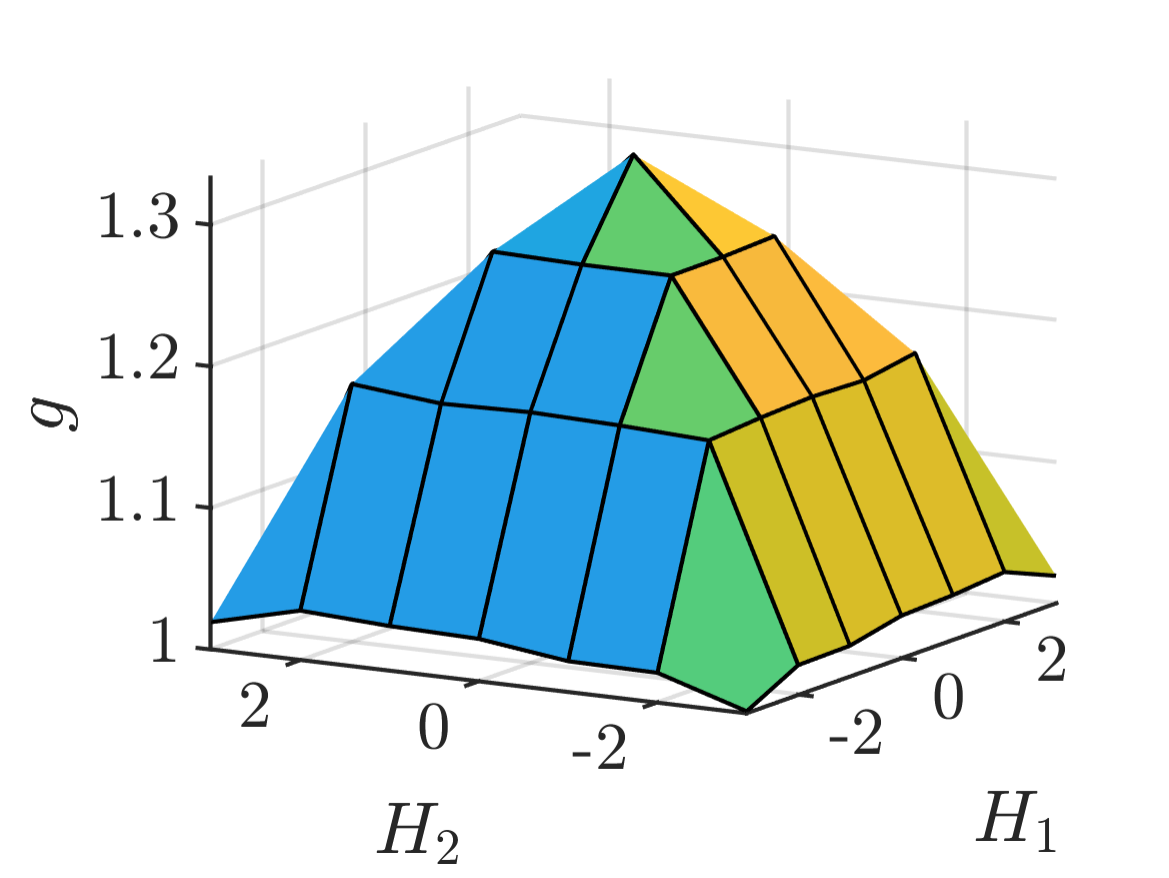}

(c)
\caption{Metrological gain for diagonal local Hamiltonians $H_1$ and $H_2$ with $+1$ and $-1$ elements, as a function of the number of $+1$'s
in $H_1$ and $H_2$. On the axes we have excess of $1$'s compared with $D/2,$ i.e., the dimension over two, in $H_k.$ Thus, $0$ corresponds to equal number of $-1$'s and $+1$'s. The gain is maximal if the number of $+1$'s is half of the dimension. That is, half of the elements are $+1,$ half of them are $-1.$ (a) $4\times4,$ (b) $6\times6,$ and (c) $8\times8$ systems.
} \label{fig:pm1}
\end{figure}

The Hamiltonian in Eq.~(\ref{eq:Ham}) can be written as in Eq.~(\ref{eq:Ham22}), where $c_1=c_2=1$ and 
\begin{equation}
\tilde D_{XX'}=(\sigma_z)_X\otimes\openone_{X'},\label{eq:DXXprime}
\end{equation}
for $X=A,B.$ Thus, the local Hamiltonians $\tilde D_{XX'}$ are diagonal and they have $\pm 1$ eigenvalues. Due to the special structure of the Hamiltonian, namely, that it acts only on $A$ and $B,$ and does not act on $A'$ and $B',$ we can write that 
\begin{equation}
(\Delta H)^2_{\varrho}=(\Delta h_{AB})^2_{\varrho_{AB}},\label{eq:redAB}
\end{equation}
where the two-qubit Hamiltonian is
\begin{equation}
h_{AB}=\sigma_z\otimes\openone+\openone\otimes\sigma_z,
\end{equation}
and $\varrho_{AB}$ is the reduced state of $A$ and $B.$ Based on Observation 2,
from Eq.~(\ref{eq:redAB})
\begin{equation}
{\cal F}_Q[\rhopptn,H]=4(\Delta h_{AB})^2_{{\rm Tr}_{A'B'}(\rhopptn)}
\end{equation}
follows. Thus, the quantum Fisher information depends only on the variance of the $h_{AB}$ in the two-qubit subsystem $AB.$

\begin{figure}[t!]
\includegraphics[width=\columnwidth]{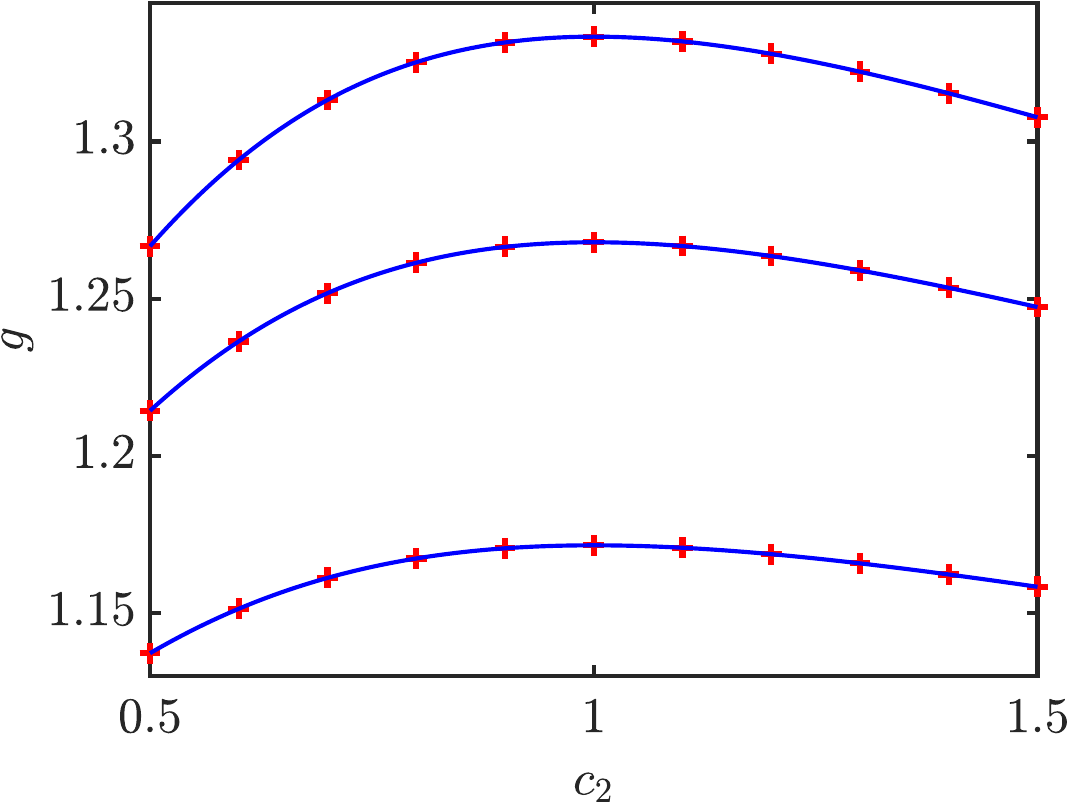}
\caption{Maximal metrological gain for a Hamiltonian of the type given in Eq.~(\ref{eq:Ham22}) if the number of $+1$'s and $-1$'s is equal in the diagonals of the local Hamiltonians and  we are changing $c_2$ while we keep $c_1=1$ for (bottom to top) $4\times4, 6\times6,$ and $8\times8$ systems. (crosses) The maximum metrological gain obtained numerically. (solid) The metrological gain for the bound entangled states based on private states, $\rhopptfirst,$ given in Eq.~(\ref{eq:rhopri}). The $u_{ij}$ are chosen based on the quantum Fourier transform. As it can be seen, the bound entangled state $\rhopptfirst$  has the largest gain among PPT quantum states even if $c_2\ne 1.$ }
\label{fig:c2}
\end{figure}

We can see that  the Hamiltonian in Eq.~(\ref{eq:Ham22}) is more general than that in Eq.~(\ref{eq:Ham}). We should now examine, which other Hamiltonian we must also consider in our search for the Hamiltonian making the largest metrological gain possible. For instance, we can examine Hamiltonians of the type Eq.~(\ref{eq:Ham22}) with $c_1=c_2=1,$ but with unequal number of $+1$'s and $-1$'s in the diagonals  of $\tilde D_{AA'}$ and $\tilde D_{BB'}.$ Note that from the point of view of the maximal gain among PPT states, only the number of the $+1$'s and $-1$'s in the diagonals of the local Hamiltonians matter, and their order does not matter.

We carried out extensive numerics for local Hamiltonians of this type. We used the method of Ref.~\cite{Toth18} that optimizes the quantum Fisher information over PPT states for a given local Hamiltonian. Figure~\ref{fig:pm1} shows that the metrological gain is the largest if the local Hamiltonians have equal number of $+1$'s and $-1$'s.

We should also examine Hamiltonians in which the local Hamiltonians are diagonal and the number of $+1$ and $-1$ eigenvalues are the same for the two subsystems, but the weights of the two local Hamiltonians are different from each other, i.e., such that $c_1=1$ and $c_2\ne 1.$ We carried out extensive numerics for such Hamiltonians. We looked for the quantum state with the largest metrological gain among PPT states for the given Hamiltonian, using the method of Ref.~\cite{Toth18}. Figure~\ref{fig:c2} shows that the maximal gain is obtained for $c_2=1.$ 

The bound entangled state $\rhopptfirst$  has the largest gain among PPT quantum states even if $c_2\ne 1,$  where the $u_{ij}$ are chosen based on the quantum Fourier transform, as can also be seen in Fig.~\ref{fig:c2}. We find that $\rhopptsecond$ is also optimal for $c_2=1,$ while it is not optimal any more if $c_2\ne1.$ The explanation for the difference is the following. The vectors appearing in Eq.~(\ref{eq:eig}) remain eigenvectors of the general Hamiltonian for $c_2\neq 1$, but the eigenvalues change. There are no zero eigenvalues any more, and $|01ij\rangle$ and $|10ij\rangle$ correspond to different eigenvalues. Therefore, states living in the subspace spanned by $|01ij\rangle$ and $|10ij\rangle$ may contribute to the quantum Fisher information. However, due to the $(\lambda_{\mu}-\lambda_{\nu})^2$ factor in Eq.~(\ref{eq:FQ}), simultaneous eigenvectors of the density matrix and the Hamiltonian still do not contribute. All eigenstates of $\rhopptsecond$ living in the space of $|01ij\rangle$ and $|10ij\rangle$  are like that, while for $\rho_{F1}$ eigenvectors $|w_i\rangle$ and $|w_i^-\rangle$ are not eigenstates any more of the modified Hamiltonian, and they together give an extra nonzero contribution.

Finally, based on the discussion above, we carried out an optimization over local Hamiltonians of the type in Eq.~(\ref{eq:Ham22}). We were optimizing over the number of $+1$'s and $-1$'s in the diagonal of the local Hamiltonians for a range of $c_2$ values and for $4\times4, 6\times6$ and $8\times8$ systems. We found that even if $c_2\ne1$ the optimum is reached by local Hamiltonians which have equal number of $+1$'s and $-1$'s in the diagonal. We used the method of Ref.~\cite{Toth18} optimizing over PPT states for a given Hamiltonian. We could not reach a larger metrological gain than with the $\rhopptfirst$ and $\rhopptsecond$ states.

\section{Discussion}\label{sec:disc}

We now discuss important properties of the quantum states presented.

\subsection{Extremal PPT states}

We mention a relation of maximally useful PPT states to extremal PPT states. At least one of the quantum states maximizing the  quantum Fisher information over the set of PPT states must be an extremal state, see Fig.~\ref{fig:pptextremal}. It can also happen that such a state is a mixture of extremal PPT states, which all have maximal quantum Fisher information.  If one encounters such a quantum state, one can try to obtain the extremal components with maximal quantum Fisher information from them. Extremal PPT states have been studied a lot \cite{Augusiak2010Searching,Oyvind2013Extremal,Leinaas2007Extreme,Leinaas2010Numerical,Leinaas2010Low}. A related notion is edge states \cite{Kraus2000Separability,Sanpera2001Schmidt}.

Indeed, our first example, $\rhopptfirst$ defined in Eq.~(\ref{eq:rhopri}), have been extremal PPT states \cite{Badziag}.  We now discuss whether $\rhopptsecond$ are also extremal PPT states. Based on  Eq.~(\ref{eq:rhorank2}), the rank of the state found for $d=2$ is $8.$  Based on Eq.~(\ref{eq:rhorank2pt}), the rank of partial transpose is $10.$ This is allowed by the rank-constraints for extremal PPT states \cite{Leinaas2010Numerical}.

\subsection{Global optimality among PPT states}

{\it Arguments for Conjecture 1.}---We showed that two different families of PPT entangled states perform the same way metrologically. Surprisingly, they approach the maximal metrological performance for large dimensions. The first family of states were extremal \cite{hllo}. These states are PPT states that are nearly as far from separable states as possible \cite{hllo}. The second family of states coincides with the states found for $d=3,4,5,$ and $6$  in Ref.~\cite{Toth18} based on extensive numerical optimization, see Appendix~\ref{sec:mapping}. The two families, found independently, have the same quantum Fisher information. Hence, we conjecture that these states are optimal for metrology among PPT states. Thus, they are likely the best we can have, if only PPT states are allowed.  

We know that local operations and classical communication (LOCC) conserve the PPT property, thus from PPT states we can never get non-PPT states. Based on this observation, one can imagine a scenario where we distill other, less useful PPT states into the ones presented in this paper. Thus, these states can play the role of ``PPT singlets.''

Let us now examine, how such a distillation scheme can look. The distillation protocol cannot be based  on depolarizing the state into isotropic states as in Ref.~\cite{distill}, because isotropic states cannot be PPT entangled. A simple distillation protocol can be based on the following idea. Let us assume that many copies of the quantum state are available. First, we make a quantum state tomography using some of the copies. Then, based on the tomography, we find the local filter that increases the quantum Fisher information the most \cite{Verstraete2001Local}. Using a positive operator-valued measure (POVM) that realizes this optimal local filter, we obtain a smaller quantity of states, but with a larger quantum Fisher information.  Local filtering has already been used to transform PPT states \cite{Tendick2020Activation}.

We show now a concrete example. Let us consider the quantum state
\begin{equation}
\varrho_{\rm noisy}=(1-p_{\rm n})\rhopptfirst+p_{\rm n}\varrho_{{\rm noise},AA'}\otimes\varrho_{{\rm noise},BB'},
\end{equation}
where $\rhopptfirst$ is the $4\times 4$ state given in Eq.~(\ref{eq:rhopri}),  the noise fraction is $p_{\rm n}=0.092,$ and the local noise is defined as
\begin{equation}
\varrho_{\rm noise}=\frac{\ketbra{00}{00}+\ketbra{11}{11}}{2}.
\end{equation}
For the noisy state we have
\begin{equation}
{\cal F}_Q[\varrho_{\rm noisy},H]=7.8828,
\end{equation}
which is smaller than the maximum for separable states ${\cal F}_Q^{\rm (sep)}(H)=8,$ see Eq.~(\ref{eq:fqsepH}).
Then, let us define the operator
\begin{equation}
F=\openone-0.1(\ketbra{00}{00}+\ketbra{11}{11}).
\end{equation}
Using $F$ as a local filter, let us generate the following quantum state
\begin{equation}
\varrho_{\rm filtered}=\frac{(F\otimes F)\varrho_{\rm noisy}(F\otimes F)^\dagger}{{\rm Tr}[(F\otimes F)\varrho_{\rm noisy}(F\otimes F)^\dagger]}.
\end{equation}
For the filtered state we have
\begin{equation}
{\cal F}_Q[\varrho_{\rm filtered},H]=8.0084,
\end{equation}
which is larger than the maximum for separable states, see Eq.~(\ref{eq:fqsepH}). Thus, the noisy state was not useful for metrology with the Hamiltonian $H$ given in Eq.~(\ref{eq:Ham}), but with local filtering it became useful. With extensive numerics using the method for optimizing the Hamiltonian for a given quantum state in Ref.~\cite{Toth2020Activating}, we find that $\varrho_{\rm noisy}$ is not useful metrologically with any other Hamiltonian.

\section{Conclusions}\label{sec:conc}

We studied metrologically useful PPT bound entangled states with large metrological gain with respect to separable states. We considered two families of ($2d\times 2d$)-dimensional PPT states. The first construction is the one given by Badzi{\k{a}}g {\it et al.}~\cite{Badziag}, whereas the second construction seem to be the state obtained from the algorithm given in Ref.~\cite{Toth18}. In both cases we calculate the metrological advantage and the robustness of metrological usefulness to noise and show that the metrological advantage monotonically increases with increasing $d.$ When $d$ goes to infinity, the state becomes maximally useful. The robustness values calculated in Sec.~\ref{sec:QFIppt} indicate that some of the presented PPT states might be implemented in the laboratory since they are robust to the noise level present in current experiments \cite{experiment}. The MATLAB routines defining the states of this paper are listed in Appendix~\ref{sec:MATLAB}.

\section{Acknowledgments} 

We thank R.~Augusiak, O.~G\"uhne, P.~Horodecki, R.~Horodecki, J.~Ko\l ody\'nski, A.~Rutkowski, and J.~Siewert for discussions. We acknowledge the support of the  EU (COST Action CA15220, QuantERA CEBBEC, QuantERA eDICT), the Spanish MCIU (Grant No. PCI2018-092896), the Spanish Ministry of Science, Innovation and Universities and the European Regional Development Fund FEDER through Grant No. PGC2018-101355-B-I00 (MCIU/AEI/FEDER, EU), the Basque Government (Grant No. IT986-16), and the National Research, Development and Innovation Office NKFIH (Grant No.  K124351, No. KH129601, No. KH125096,  and No. 2019-2.1.7-ERA-NET-2020-00003).  We thank the ``Frontline'' Research Excellence Programme of the NKFIH (Grant No. KKP133827). G.T. is grateful for a  Bessel Research Award of the Humboldt Foundation.

\appendix

\section{Family of PPT states based on private states}
\label{sec:privatePPT}

In this Appendix, we briefly review private states~\cite{hllo} (see e.g.,~Ref.~\cite{hhho} for more details). Private states are quantum states shared among four systems $A, A', B,$ and $B'$. The systems $A,B$ form the key part, whereas $A',B'$ belong to the shield part. Alice's component spaces are $\mathcal{H}_A\otimes\mathcal{H}_{A'}$ and Bob's component spaces are $\mathcal{H}_B\otimes\mathcal{H}_{B'}$, respectively. We focus on $\mathcal{H}_A=\mathcal{H}_B=\CC^2$ and $\mathcal{H}_A'=\mathcal{H}_B'=\CC^d$, in which case we call the private state a private bit. 

A generic private bit has the following form~\cite{hllo,hhho}
\begin{equation}
\varrho_{\rm bit}\equiv U\ketbra{\phi_+}{\phi_+}\otimes \sigma_{A'B'}U^{\dagger},
\end{equation}
where the maximally entangled two-qubit state is defined as
\begin{equation}
\ket{\phi_+}=(\ket{00}+\ket{11})/\sqrt 2, \label{eq:maxent}
\end{equation}
$\sigma_{A'B'}$ is a state of systems $A'B'$, and $U$ is an arbitrary twisting operation which can be written in the following form~\cite{hhho}
\begin{equation}
U=\sum_{k,l=0}^1{\ketbra{kl}{kl}}_{AB}\otimes U_{A'B'}^{kl}.
\end{equation}

It turns out that any private bit can be written in the Hilbert space $\mathcal{H}_A\otimes\mathcal{H}_{B}\otimes\mathcal{H}_{A'}\otimes\mathcal{H}_{B'}$ in the form
 \begin{equation}
\varrho_{\rm bit}=\varrho(X)=\frac{1}{2}
\left[{\begin{array}{cccc}
\sqrt{XX^\dag}&0&0&X\\
0&0&0&0\\
0&0&0&0\\
X^\dag&0&0&\sqrt{X^\dag X}\\
\end{array} } \right],
\label{eq:rhobit}
\end{equation}
up to a change of basis in the key part $AB$, where $X$ is a trace norm 1 operator. 

Consider now two matrices with a unit trace norm given in Eq.~(\ref{eq:XY}). The family of PPT states constructed in Ref.~\cite{Badziag} is a mixture of two mutually orthogonal private bits and looks as
\begin{equation}
p_1\varrho(X)+p_2\varrho',
\label{eq:rhoPPT}
\end{equation}
where the density matrix $\varrho'$ is defined as
\begin{equation}
\varrho'=\sigma_A^x\otimes  \openone_{BA'B'}\varrho(Y)\sigma_A^x\otimes  \openone_{BA'B'},
\end{equation}
where $\sigma_A^x$ is a Pauli matrix, and the weights are given in Eqs.~(\ref{eq:p1def}) and (\ref{eq:p2def}).
Note that $\varrho(X)$ and $\varrho'$ appearing in  Eq.~(\ref{eq:rhoPPT}) are two mutually orthogonal private bits. 

In matrix notation, the states $\rhopptfirst$ given in Eq.~(\ref{eq:rhoPPT}) can be written as in Eq.~(\ref{eq:rhomat}). Using Eqs.~(\ref{eq:XY}) and (\ref{eq:XXYY}), and complementing the vectors with the indices corresponding to spaces $\mathcal{H}_A$ and $\mathcal{H}_B$, the density matrix in Eq.~(\ref{eq:rhomat}) can be expressed in Hilbert space $\mathcal{H}_A\otimes\mathcal{H}_{B}\otimes\mathcal{H}_{A'}\otimes\mathcal{H}_{B'}$ as in Eq.~(\ref{eq:rhopri}).

\section{Transforming the $4\times4$ bound entangled state presented in Ref.~\cite{Toth18} to a private state}
\label{sec:appendix_bound_private}

In this Appendix, we show that the $4\times4$ state presented in Ref.~\cite{Toth18} can be mapped to a private state with local unitary transformations and relabeling of the subsystems within the two subsystems. Private states have been presented in  Ref.~\cite{hllo} and we also reviewed them in Appendix~\ref{sec:privatePPT}.

The  $4\times 4$ bound entangled PPT state given in Ref.~\cite{Toth18}  is the following.  The system is partitioned into subsystems $A$ and $B,$  where both $A$ and $B$ are four dimensional. Let us define the following six states 
\begin{align}
\ket{\Psi_1}&=(\ket{0,1}+\ket{2,3})/\sqrt 2,\nonumber\\
\ket{\Psi_2}&=(\ket{1,0}+\ket{3,2})/\sqrt 2,\nonumber\\
\ket{\Psi_3}&=(\ket{1,1}+\ket{2,2})/\sqrt 2,\nonumber\\
\ket{\Psi_4}&=(\ket{0,0}-\ket{3,3})/\sqrt 2,\nonumber\\
\ket{\Psi_5}&=(+\ket{0,3}+\ket{1,2})/2+\ket{2,1}/\sqrt 2,\nonumber\\
\ket{\Psi_6}&=(-\ket{0,3}+\ket{1,2})/2+\ket{3,0}/\sqrt 2.
\end{align}
Then the  state is given as
\begin{align}
\varrho_{4\times4}= p\sum_{n=1}^4\ket{\Psi_n}\bra{\Psi_n}+q\sum_{n=5}^6\ket{\Psi_n}\bra{\Psi_n},
\label{eq:rho4x4}
\end{align}
where the eigenvalues of the eigendecomposition are 
\begin{align}
q&=(\sqrt 2 - 1)/2,\nonumber\\
p&=(1 - 2q)/4. \label{eq:eigrho4x4}
\end{align}
Note that the state given in Eq.~(\ref{eq:rho4x4}) is permutationally invariant, that is,
\begin{equation}
F \varrho_{4\times4} F = \varrho_{4\times4},
\end{equation}
where $F$ is the flip operator.

Let us see now the metrological performance of the state given in Eq.~(\ref{eq:rho4x4}). We consider the operator 
\begin{align}
H_{4\times4}&=D\otimes\openone+\openone\otimes D,
\end{align}
 where $D={\rm diag}(1,1,-1,-1)$. 
Then, we obtain 
\begin{align}
\mathcal{F}_Q[\varrho_{4\times4},H_{4\times4}]=32-16\sqrt 2\simeq9.3726,
\end{align}
where for separable states the maximum of the quantum Fisher information is $8.$

Let us see now the $4\times4$ private state of Ref.~\cite{hllo}, for which $d=2.$ We use the formula Eq.~(\ref{eq:rhopri}), where the eigenvectors with nonzero eigenvalue are given in  Eqs.~(\ref{eq:vkl}) and (\ref{eq:wk}). We choose
\begin{equation}
u=\frac 1 {\sqrt2}\left[\begin{array}{rr} 1&1\\1&-1\\ \end{array}\right].\label{eq:u1m1}
\end{equation}
Based on that, we obtain
\begin{align}
\rhopptfirst=\Lambda_v \sum_{i,j=0,1} |v_{ij}\rangle\langle v_{ij}|+\Lambda_w \sum_{i=0,1} |w_{i}\rangle\langle w_{i}|,
\label{eq:rhoppt1}
\end{align}
where, based on Eqs.~(\ref{eq:lambdav}) and (\ref{eq:lambdaw}) the eigenvalues are 
\begin{align}
\Lambda_v&=(\sqrt2-1)/(2\sqrt2),\nonumber\\
\Lambda_w&=(\sqrt 2 - 1)/2
\label{eq:eigrhoppt1}
\end{align}
and, based on Eqs.~(\ref{eq:vkl}) and (\ref{eq:wk}), the eigenvectors are 
\begin{align}
\ket{v_{00}}&=(\ket{0000}+\ket{1100})/\sqrt 2,\nonumber\\
\ket{v_{01}}&=(\ket{0001}+\ket{1110})/\sqrt 2,\nonumber\\
\ket{v_{10}}&=(\ket{0010}+\ket{1101})/\sqrt 2,\nonumber\\
\ket{v_{11}}&=(\ket{0011}-\ket{1111})/\sqrt 2,\nonumber\\
\ket{w_0}&=\ket{0100}/\sqrt 2+ (\ket{1000}+\ket{1011})/2,\nonumber\\
\ket{w_1}&=\ket{0111}/\sqrt 2+ (\ket{1000}-\ket{1011})/2,\label{eq:eigrhoppt1v}
\end{align}
where the order of the subsystems of the four-qubit system is $ABA'B'.$

We now show how to map the private state given in Eq.~(\ref{eq:rhoppt1}) to the state given in Eq.~(\ref{eq:rho4x4}). First, we carry out a transformation from a two-qubit system living on $AA'$ to a four-dimensional system in $B$ as
\begin{align}
|00\rangle &\rightarrow |1\rangle,\nonumber\\
|01\rangle &\rightarrow |0\rangle,\nonumber\\
|10\rangle &\rightarrow |2\rangle,\nonumber\\
|11\rangle &\rightarrow |3\rangle, \label{eq:transf}
\end{align}
Similarly, on the space of Bob, we carry out a transformation from a two-qubit system living on $BB'$ to a four-dimensional system in $A.$ Using the transformation above, the two states correspond to each other.

In detail, we can map the eigenvectors to each other as
\begin{align}
v_{00} &\rightarrow |\Psi_3\rangle,\nonumber\\
v_{01} &\rightarrow |\Psi_1\rangle,\nonumber\\
v_{10} &\rightarrow |\Psi_2\rangle,\nonumber\\
v_{11} &\rightarrow |\Psi_4\rangle,\nonumber\\
w_0 &\rightarrow |\Psi_5\rangle,\nonumber\\
w_1 &\rightarrow |\Psi_6\rangle,
\end{align}
while the eigenvalues given in Eqs.~(\ref{eq:eigrho4x4}) and (\ref{eq:eigrhoppt1}) are identical.

\section{An alternative of the algorithm maximizing the quantum Fisher information in Ref.~\cite{Toth18} }
\label{sec:alternative}

\newcommand{\va}[1]{\ensuremath{(\Delta#1)^2}}
\newcommand{\vasq}[1]{\ensuremath{[\Delta#1]^2}}
\newcommand{\varho}[1]{\ensuremath{(\Delta_\rho #1)^2}}
\newcommand{\ex}[1]{\ensuremath{\langle{#1}\rangle}}
\newcommand{\exs}[1]{\ensuremath{\langle{#1}\rangle}}

In Ref.~\cite{Toth18}, a method is presented that maximizes the quantum Fisher information over density matrices for a given Hamiltonian, which used semidefinite programming. One can combine this with the approach of Refs.~\cite{Macieszczak1,Macieszczak2,Chabuda2020}
as follows.
We write the optimization of the quantum Fisher information  over the state as an optimization of another quantity, the error propagation formula
\begin{equation}
\label{eq:vartheta}
\va{\theta}_M= \frac{\va{M}}{\ex{{\bf i}[M,{ H}]}^2},
\end{equation}
over the state $\varrho$ and $M$ as \cite{Macieszczak1,Macieszczak2}
\begin{eqnarray}
&&\max_{\varrho\in\mathcal S}F_Q[\varrho,H]\nonumber\\
&&\quad\quad=\max_{\varrho\in\mathcal S} \max_{M} {1/\va{\theta}_M }\nonumber\\
&&\quad\quad=\max_{\varrho\in\mathcal S} \max_{M} {\ex{{\bf i}[M,{H}]}^2}/{\ex{M^2} }\nonumber\\
&&\quad\quad=\max_{\varrho\in\mathcal S} \max_M \max_\alpha \{ -\alpha^2 \ex{M^2} + 2 \alpha \ex{{\bf i}[M,{ H}]}  \}\nonumber\\
&&\quad\quad=\max_{\varrho\in\mathcal S} \max_{M'} \{ -\ex{(M')^2} + 2 \ex{{\bf i}[M',{ H}]} \},\label{eq:fmin}
\end{eqnarray}
where $M'$ takes the role of $\alpha M$ and $\mathcal S$  denotes the set of PPT states. 

For a constant $M'$, the maximization over $\varrho$ for PPT states can be carried out by semidefinite programming similarly as in Ref.~\cite{Toth18}, since the expression to be optimized is linear in $\varrho.$
For a given state $\varrho$, the maximization over $M'$ can be carried out by setting $M'$ to be the symmetric logarithmic derivative defined as
\begin{equation}
\label{eq:SLD}
M_{\rm opt}=2{\bf i}\sum_{k,l}\frac{\lambda_{k}-\lambda_{l}}{\lambda_{k}+\lambda_{l}} \vert k \rangle \langle l \vert \langle k \vert {\mathcal H} \vert l \rangle.
\end{equation}
By alternating the optimization over $\varrho$ and $M',$ we can reach the optimal $\varrho.$

The approach presented in this section might be somewhat faster than the approach given in Ref.~\cite{Toth18} in some situations. On the other hand, the method in Ref.~\cite{Toth18} has the advantage that it optimizes the error propagation formula directly and provides its value at each step of the iteration \cite{thanks}.

\section{Algorithm maximizing the quantum Fisher information in Ref.~\cite{Toth18} finds states in the $\rhopptsecond$ family}
\label{sec:mapping}

In Ref.~\cite{Toth18}, a method is presented that maximizes the quantum Fisher information over density matrices for a given Hamiltonian. For $d=3,4,5,$ and $6,$ concrete numerical examples are presented where the Hamiltonian is identical to Eq.~(\ref{eq:Ham}), apart from a trivial rearrangement of the parties. The eigenvalues of the density matrix and its partial transpose equal to those of the corresponding states in the $\rhopptsecond$ family. Next, we will argue that the states are within the $\rhopptsecond$ family.

Let us consider a local Hamiltonian $\mathcal K$ which commutes with $\Hop$ 
\begin{equation}
[\Hop,\mathcal K]=0.\label{eq:commKH}
\end{equation}
Let us define the unitary evolution 
\begin{equation}
U_t=\exp(-{\bf i}\mathcal K t).\label{eq:UKT}
\end{equation}
For such an evolution we have 
\begin{equation}
\mathcal{F}_Q[U_t\varrho U_t^\dagger,\Hop]=
\mathcal{F}_Q[\varrho ,U_t^\dagger \Hop U_t]=\mathcal{F}_Q[\varrho,\Hop].\label{eq:FQQ}
\end{equation}
In Eq.~(\ref{eq:FQQ}), the first equality can be directly proved based on Eq.~(\ref{eq:FQ}). The second equality is the result of Eq.~(\ref{eq:commKH}). Thus, the quantum Fisher information does not change in an evolution given in Eq.~(\ref{eq:UKT}), if Eq.~(\ref{eq:commKH}) is fulfilled.

One can consider the relevant case, when $\mathcal H$ is defined as in Eq.~(\ref{eq:H12}), where $\Hop_k$ are local operators. We would define $\mathcal K$ similarly as 
\begin{equation}
\mathcal K=\mathcal K_1+\mathcal K_2 \label{eq:K12}
\end{equation}
where $\mathcal K_1$ and $\mathcal K_2$ are single-particle operators acting on the two subsystems. Then, if
\begin{equation}
[\Hop_1,\mathcal K_1]=0,\quad [\Hop_2,\mathcal K_2]=0\label{eq:Kcond}
\end{equation}
is fulfilled then a unitary evolution given in Eq.~(\ref{eq:UKT}) does not change the quantum Fisher information. Moreover, during dynamics under $\mathcal K$ given in Eq.~(\ref{eq:K12}), separable states remain separable and PPT states remain PPT.

Based on this, we can see that an algorithm maximizing the quantum Fisher information over quantum states typically will not find a unique solution. On the other hand, we can try to transform the quantum states found numerically to the analytic families given in this paper. Assume that we want to show that  $\varrho_A$ can be transformed to $\varrho_B.$ The algorithm is the following.

(i) We set $\varrho=\varrho_A.$   (ii) We generate random $\mathcal K_1$ and $\mathcal K_2$ fulfilling Eq.~\eqref{eq:Kcond}, and compute 
\begin{equation}
\varrho'=\exp(-{\bf i}\mathcal KT)\varrho\exp(+{\bf i}\mathcal KT),
\end{equation}
where $T$ is some constant. (iii) If 
\begin{equation}
||\varrho'-\varrho_B||<||\varrho-\varrho_B||,
\end{equation}
where $||A||$ is a norm, then we set $\varrho=\varrho'.$ (iv) We go back to step (ii).

We find the $4\times 4$ states found by the algorithm in Ref.~\cite{Toth18}, can be transformed to the $4\times 4$ members of the $\rhopptfirst$ family or those of the $\rhopptsecond$ family. In particular, if we iterate the search algorithm for a few steps, it might find a state equivalent to $\rhopptfirst$. If we iterate further, it finds a state equivalent to $\rhopptsecond,$ which has a higher rank. We could also show that the $6\times 6$  states found numerically can be converted to the $6\times 6$ member of the $\rhopptsecond$ family. 

For larger system sizes, such a search is difficult to carry out. Even in this case, based on extensive numerics, we can see that  the states in the $\rhopptsecond$ family are fixed points of the  algorithm maximizing the quantum Fisher information presented in Ref.~\cite{Toth18}. If we start out from these states, the algorithm will not change the state. If we disturb the state by a small amount of added noise, the algorithm will end up in a state that is close to the original state. On the other hand, if we start out from a state in the $\rhopptfirst$ family, then the algorithm will leave the state and end up in a state  of the $\rhopptsecond$ family. The algorithm prefers $\rhopptsecond$ possibly as it has a higher rank than $\rhopptfirst.$ Additional constraints (e.g., rank-constraints) might be used to force the algorithm to settle in states of the $\rhopptfirst$  family. In Sec.~\ref{sec:GenHam}, we find that there are Hamiltonians for which $\rhopptfirst$ is optimal from the point of view of metrological gain, while  $\rhopptsecond$ is not optimal.

\section{Symmetry properties of the $4\times 4$ PPT state based on private states}
\label{sec:appendix4by4state}

In this Appendix, we prove that for $d=2$ the state $\rhopptfirst$ in Eq.~(\ref{eq:rhopri}) can be written as an equal mixture of symmetric and antisymmetric states. That is, the state can be written as
\begin{equation}
\rhopptfirst=(1/2)\varrho_{\cal S} + (1/2)\varrho_{\cal A}. 
\label{eq:rhoas}
\end{equation}
The state $\varrho_{\cal A}$ acts only on the antisymmetric subspace $\cal{H_A}$, whereas $\varrho_{\cal S}$ acts only on the symmetric subspace $\cal{H_S}$. 
We also show how to choose $u_{ij}$ such that the state remains the mixture of a symmetric and antisymmetric part. 

In the case of $d=2$, the eigenvalues are listed in Eq.~(\ref{eq:eigrhoppt1}). The first eigenvalue is four-times degenerate, the second one is twice degenerate. The corresponding eigenvectors are listed in Eq.~(\ref{eq:eigrhoppt1v}). The first four eigenvectors belong to the first eigenvalue, the last two eigenvectors belong to the second eigenvalue.

Below it is shown that by choosing different eigenstates which span the same eigenspace for the respective eigenvalues 
each eigenvector will take the form of either a symmetric or an antisymmetric state. To this end, let us define the new eigenstates
\begin{align}
\ket{v_{01}'}&=(\ket{v_{01}}+\ket{v_{10}})/\sqrt 2,\nonumber\\
\ket{v_{10}'}&=(\ket{v_{01}}-\ket{v_{10}})/\sqrt 2
\end{align}
 to get a symmetric and an antisymmetric state as follows
\begin{align}
\ket{v_{01}'}&=(\ket{00}\ket{01}+\ket{01}\ket{00}+\ket{11}\ket{10}+\ket{10}\ket{11})/2, \nonumber\\
\ket{v_{10}'}&=(\ket{00}\ket{01}-\ket{01}\ket{00}+\ket{11}\ket{10}-\ket{10}\ket{11})/2,
\end{align}
where we shorthanded $\ket{\alpha\beta}\ket{\gamma\delta}\equiv\ket{\alpha\beta}_{AA'}\ket{\gamma\delta}_{BB'}$. Similarly, we define 
\begin{align}
\ket{w_{0}'}=(\ket{w_{0}}+t\ket{w_{1}})/\sqrt {1+t^2},\nonumber\\
\ket{w_{1}'}=(\ket{w_{1}}-t\ket{w_{0}})/\sqrt {1+t^2}
\end{align}
to get the respective symmetric and antisymmetric states:
\begin{align}
\ket{w_{0}'}&=c_+(\ket{00}\ket{10}+\ket{10}\ket{00})+c_-(\ket{01}\ket{11}+\ket{11}\ket{01}), \nonumber\\
\ket{w_{1}'}&=c_+(\ket{01}\ket{11}-\ket{11}\ket{01})+c_-(\ket{10}\ket{00}-\ket{00}\ket{10}), 
\end{align}
where we have chosen 
\begin{equation}
t=\sqrt 2-1,
\end{equation}
and we write 
\begin{equation}
c_{\pm}=(1/2)\sqrt{1\pm{}1/\sqrt{2}}.
\end{equation}
Let us observe that the eigenstates $\ket{v_{00}},\ket{v_{01}'},\ket{w_0'}$ above are symmetric, whereas the eigenstates $\ket{v_{11}}$,$\ket{v_{10}'},\ket{w_1'}$ above are antisymmetric. This implies that $\rhopptfirst$ can be written 
as an equal mixture of $\varrho_{\cal S}$ and $\varrho_{\cal A}$, that is in the form~(\ref{eq:rhoas}), where the density matrices acting on the respective symmetric and antisymmetric spaces are 
\begin{align}
\varrho_{\cal S}&=  (p_1/2)\ket{v_{00}}\bra{v_{00}}+(p_1/2)\ket{v_{01}'}\bra{v_{01}'}+p_2\ket{w_{0}'}\bra{w_{0}'} ,\nonumber\\
\varrho_{\cal A}&=  (p_1/2)\ket{v_{11}}\bra{v_{11}}+(p_1/2)\ket{v_{10}'}\bra{v_{10}'}+p_2\ket{w_{1}'}\bra{w_{1}'}.
\end{align}

Let us see the case of $d>2.$ We will show that it is still possible to write the state as a mixture of a symmetric and antisymmetric state, if we choose $u$ properly. In case of a bipartite state, this is equivalent to that the state is permutationally invariant, that is
\begin{equation}
F \rhopptfirst F^\dagger = \rhopptfirst,
\end{equation}
where $F$ exchanges $A$ and $B$, and exchanges $A'$ and $B'.$

{\bf Observation 6.}---The bound entangled state $\rhopptfirst$ based on private states given in Eq.~(\ref{eq:u1m1}) is permutationally invariant, if and only if $u_{ij}$ is a real and symmetric matrix,
where for each element $u_{ij}=\pm1/\sqrt{d}.$ If $d$ is a power of two, i.e., $d=2^n,$ where $n$ is an integer, we can choose $u$ to be
\begin{equation}
u=2^{-n/2}\left[\begin{array}{rr} 1&1\\1&-1\\ \end{array}\right]^{\otimes n}.\label{eq:u1m1b}
\end{equation}

{\it Proof.} To see this, let us consider the formula in Eq.~(\ref{eq:rhopri}) defining $\rhopptfirst.$ The first and third sums are permutationally invariant. The second sum is permutationally invariant if
\begin{align}
u_{ij}=u_{ji}.
\end{align}
The fourth sum is permutationally invariant if
\begin{align}
u_{ij}=u_{ij}^*.
\end{align}
One can see that the expression is permutationally invariant, if the 
four sums are  permutationally invariants.
Hence, Eq.~(\ref{eq:rhopri}) is permutationally invariant if and only if $u_{ij}$ is a real symmetric matrix. Moreover, due to Eq.~(\ref{eq:uij}),
all $u_{ij}$ must be $\pm 1/\sqrt d.$

For $d=2,$ where $u_{ij}$ is given in Eq.~(\ref{eq:u1m1}), the requirements  above are fulfilled.
For larger dimensions, if  $u_{ij}$ is the quantum Fourier transform then it does not fulfill this condition.
On the other hand, if $d$ is a power of two, i.e., $d=2^n,$ where $n$ is an integer, we can choose $u$ to be the value given in Eq.~(\ref{eq:u1m1b}). $\qed$

\section{Maximum value of the quantum Fisher information for $H$}
\label{sec:appendix0}

In this Appendix, we determine the maximal quantum Fisher information for our problems.

{\bf Observation 7.}---The maximum of the quantum Fisher information for $H$ given in Eq.~(\ref{eq:Ham}) is 16.

{\it Proof.}---We will use the series of inequalities
\begin{equation}
{\cal F}_Q[\varrho,{H}]\le 4 (\Delta H)^2_\varrho \le 4 \langle H^2 \rangle_\varrho \le 4 \lambda_{\max } (H^2) =16,\label{eq:series}
\end{equation}
where $\lambda_{\max } (A)$ is the largest eigenvalue of the matrix $A.$
In Eq.~(\ref{eq:series}), the first inequality is generally true for the quantum Fisher information and the variance, see Eq.~(\ref{eq:FQ4VAR_general}). The second inequality is a property of the variance, which is defined as
\begin{equation}
(\Delta H)^2_\varrho=\langle H^2 \rangle_\varrho-\langle H \rangle_\varrho^2.
\end{equation}
In Eq.~(\ref{eq:series}), the third inequality is based on that $\langle A \rangle \le \lambda_{\max } (A)$  holds for any operator $A.$
For the Hamiltonian in Eq.~(\ref{eq:Ham}) we have
\begin{equation}
\lambda_{\max } (H^2)=4.
\end{equation}
A state saturating all inequalities in Eq.~(\ref{eq:series}) is 
\begin{equation}
(\vert v_2 \rangle + \vert v_{-2} \rangle)/\sqrt{2},
\end{equation}
where $\vert v_\lambda \rangle$ denotes an eigenvector of $H$ with eigenvalue $\lambda.$
In the main text, the eigenvectors and eigenvalues of $H$ are given in Eq.~(\ref{eq:eig}).  $\qed$

It is easy to see that for a pair of maximally entangled  qubits living on $AB$
\begin{equation}
\varrho_{\rm me}=\vert {\phi_+} \rangle_{AB} \langle {\phi_+} \vert  \otimes \openone_{A'B'},
\end{equation}
where $\ket{\phi_+}$ is defined in Eq.~(\ref{eq:maxent}), we obtain ${\cal F}_Q[\varrho_{\rm me},{H}]=16,$
where $H$ is given in Eq.~(\ref{eq:Ham}). Thus, even a two-qubit maximally entangled state can reach the maximum, however, no higher dimensional state can surpass it.

\section{Special properties of orthogonal matrices $Q_{ik}^j$}
\label{sec:appendix_special}

In this Appendix, we will prove that the orthogonal matrices of  Eq.~(\ref{eq:zvec}) we propose for $d=3$ and $2^n$ have all the properties required to ensure that the quantum Fisher information derived in the main text is correct. We analyze the two distinct cases $d=3$ and $2^n$ in the following subsections.
\subsection{The $d=3$ case}\label{sec:d3case}
Let the orthogonal matrices of Eq.~(\ref{eq:zvec}) be
\begin{equation}
Q^k=\left({\begin{array}{ccc}
	\cos\varphi_k&\hphantom{-}\sin\varphi_k&0\\
	\sin\varphi_k&-\cos\varphi_k&0\\
	0&0&1\\
	\end{array} } \right ),
\label{eq:qk3}
\end{equation}
where $\varphi_k=2\pi k/3+\varphi_0$. From Eq.~(\ref{eq:qvec}), it follows that 
\begin{equation}
|q_{02}\rangle=|q_{12}\rangle=|q_{20}\rangle=|q_{21}\rangle=0,
\end{equation}
where we defined the $|q_{ij}\rangle$ vectors in Eq.~(\ref{eq:qvec}). The rest of the $|q_{ij}\rangle$ vectors can be arranged into three subgroups, with each group represented by one of its normalized members $\vert\bar q_m\rangle$ described around Eq.~(\ref{eq:qm}) as follows
\begin{align}
&|q_{00}\rangle=-|q_{11}\rangle=\sum_k\cos\varphi_k|10kk\rangle=\sqrt{\frac{3}{2}}|\bar q_0\rangle,\nonumber\\
&|q_{01}\rangle=\hphantom{-}|q_{10}\rangle=\sum_k\sin\varphi_k|10kk\rangle=\sqrt{\frac{3}{2}}|\bar q_1\rangle,\nonumber\\
&|q_{22}\rangle=\sum_k|10kk\rangle|=\sqrt{3}\vert\bar q_2\rangle.
\label{eq:sg3}
\end{align}
It is easy to check that with this choice of angles $\varphi_k$ the vectors $\vert\bar q_m\rangle$ are normalized and orthogonal to each other. The $S_{ij}$ factors connecting $\vert q_{ij}\rangle$ and $\vert\bar q_m\rangle$ are also as required.

From Eq.~(\ref{eq:sg3}) and from the definition of the $|t_m\rangle$ in Eq.~(\ref{eq:vectm}) it follows that 
\begin{eqnarray}
|t_0\rangle&=&(|0100\rangle-|0111\rangle)/\sqrt{2},\nonumber\\
|t_1\rangle&=&(|0101\rangle+|0110\rangle)/\sqrt{2},\nonumber\\
|t_2\rangle&=&|0122\rangle. 
\end{eqnarray}
With the choice of 
\begin{eqnarray}
|s_0\rangle&=&(|0100\rangle+|0111\rangle)/\sqrt{2}, \nonumber\\
|s_1\rangle&=&(|0101\rangle-|0110\rangle)/\sqrt{2},\nonumber\\
|s_2\rangle&=&|0122\rangle
\end{eqnarray}
 it is easy to verify that Eq.~(\ref{eq:st}) is indeed satisfied.

\subsection{The $d=2^n$ case}\label{sec:devcase}
Let us define matrices $P^k(n)$ as tensor products of the $2\times 2$ identity matrix $\openone$ and the Pauli $X$ matrix.  Let us order the $P^k(n)$ matrices as follows. We define the $P^k(n)$ matrices as
\begin{equation}
P^k(n) = X^{(1-k_{n-1})} \otimes X^{(1-k_{n-2})} \otimes ... \otimes X^{(1-k_0)},\label{eq:ordering}
\end{equation}
where $0\le k\le 2^n-1$ and we use that $X^0=\openone$ and $X^1=X.$ In the formula  Eq.~(\ref{eq:ordering}), we considered $k$ as an $n$-bit binary number, and $k_i$ denotes the $i^{th}$ bit of $k.$  For example, for $n=3$ the matrices are
\begin{align}
&\openone\otimes \openone\otimes \openone,& &\openone\otimes \openone\otimes X, &
&\openone\otimes X\otimes \openone,& &\openone\otimes X\otimes X, \nonumber\\
&X\otimes \openone\otimes \openone,& &X\otimes \openone\otimes X,& &X\otimes X\otimes \openone,& &X\otimes X\otimes X.
\end{align}

The $P$-matrices can be constructed recursively by starting from 
\begin{equation}
P^0(0)=\openone_{1\times 1}\equiv 1,
\end{equation}
where $P^0(0)$ is a $1\times1$ identity matrix. We can then use the recursive formulas
\begin{align}
P^k(n+1)&=\left({\begin{array}{cc}
	P^k(n)&{\bf 0}\\
	{\bf 0}&P^k(n)\\
	\end{array}}\right ),\nonumber\\
P^{k+2^n}(n+1)&=\left({\begin{array}{cc}
	{\bf 0}&P^k(n)\\
	P^k(n)&{\bf 0}\\
	\end{array}}\right ),
\label{eq:itP}
\end{align}
where $\bf 0$ denotes $2^n\times2^n$ matrices of zero entries. From this construction, using induction it follows that there are $d=2^n$ entries $1$ in each matrix such that there is one $1$ in each row and in each column, there are no two matrices having entry $1$ on the same place, and that the zeroth row (column) of matrix $P^k$ has entry $1$ in its $k$th place, that is 
\begin{equation}
P_{0k}^{k}(n)=P_{k0}^k(n)=1\label{eq:P0k}
\end{equation}
for $k\le n.$ Note that for the rows and columns, we use indices between $0$ and $d-1,$ which makes it possible to present our arguments in a concise form. From now on we drop argument $n$ of matrices $P^k$.

Let the orthogonal matrices of Eq.~(\ref{eq:zvec}) be the $P$-matrices themselves, that is 
\begin{equation}
Q_{ij}^k=P_{ij}^{k}. 
\end{equation}
We have to remember that at each $ij$ place one and only one matrix $Q_{ij}^k$ has a $1$ entry, the others have a zero entry. Hence, there is only a single nonzero term in the sum in Eq.~(\ref{eq:qvec}), and $|q_{ij}\rangle=|10kk\rangle$ for some $k.$ Such vectors have norm one, and pairs of them are either orthogonal or equal to each other.  We can choose the concrete $Q_{ij}^k$'s such that
\begin{equation}
|\bar q_{m}\rangle=|q_{0m}\rangle=|10mm\rangle,
\end{equation}
where $|\bar q_{m}\rangle$ is described in Eq.~(\ref{eq:qm}).
This follows from $Q_{0m}^{m}=P_{0m}^m=1,$ which is due to Eq.~(\ref{eq:P0k}). Then $(i,j)\in\Xi_{m}$ if and only if $P_{ij}^m=1$, in this case $|q_{ij}\rangle=|10mm\rangle$. There are exactly $d$ such vectors, therefore, $N_{m}=d$, which is consistent with $S_{ij}=1$.

From Eq.~(\ref{eq:vectm}) and from $S_{ij}=1$ it follows that $|t_m\rangle$ is the normalized sum of $|01ij\rangle$ vectors for all $(i,j)\in\Xi_{m}$. Now it means that 
\begin{equation}
|t_m\rangle= \ket{01}_{AB} \otimes \sum_{ij} \left(P_{ij}^m|ij\rangle\right)_{A'B'}/\sqrt{d}. 
\end{equation}
In the equation above the summation has been extended to all $(i,j)$ indices, which can be done since $P^m_{ij}=0$ whenever $(i,j) \notin\Xi_{m}$.

Next, we will prove an important statement that is needed to find the $|s_m\rangle$ vectors for our quantum state.

{\bf Observation 8.}---The matrix $\sum_{m}|t_m\rangle\langle t_m|$ is invariant under partial transposition.

{\it Proof.}---Invariance of the operator 
\begin{equation}
\sum_m|t_m\rangle\langle t_m|=\ket{01} \bra{01} \otimes \sum_{mijkl}P_{ij}^mP_{kl}^m|ij\rangle\langle kl|/d
\end{equation}
 means that it is equal to its partial transpose 
 \begin{equation}
\ket{01} \bra{01} \otimes \sum_{mijkl}P_{il}^mP_{kj}^m|ij\rangle\langle kl|/d. 
 \end{equation}
 This is true if for any $\{m, j, l\}$ there exists $m'$ such that 
$P_{ij}^{m'}=P_{il}^{m},$
 $P_{il}^{m'}=P_{ij}^{m}$
for all $i$. What we will show is that there is a transformation that permutes the columns of the $P$-matrices such that in the final matrix two specified columns will be swapped, and it will also be a $P$-matrix. Let us multiply a matrix by $P^{2^\nu}$ ($0\leq \nu<n$) from the right. For $\nu=0$ the transformation will swap every second column of the matrix, for $\nu=1$ it will swap every second pairs of columns, and so on. For $\nu=n-1$ it will swap the lower half of the columns and the upper half of them. We can move any column anywhere with such transformations: if it is not in the required half, we apply the transformation with $\nu=n-1$. After this, if it is in the wrong quarter, we apply the $\nu=n-2$ transformation, and so on. Any such transformation applied to a $P$-matrix leads to another $P$-matrix (the product of any two $P$-matrices is a $P$-matrix, which follows from their definition). Furthermore, as these transformations commute, if one moves a column from position $j$ to position $l$, the same set of transformations will move the column from position $l$ to position $j$, that is it will swap those columns, which concludes the proof. $\qed$
 
Due to Observation 8,  we can take 
\begin{equation}
|s_m\rangle=|t_m\rangle.
\end{equation}

\section{MATLAB routines}
\label{sec:MATLAB}

We used MATLAB for the calculations in this paper \cite{MATLAB}. We created MATLAB routines that define the quantum states presented in this paper.  They are part of the QUBIT4MATLAB package \cite{QUBIT4MATLAB}. The routine \verb|BES_private.m| defines the private states  given in Eq.~(\ref{eq:rhopri}). For the $u_{ij}$ unitaries, the quantum Fourier transform is used. The routine \verb|BES_metro4x4.m| defines the state  given in Eq.~(\ref{eq:rho4x4}). The routine \verb|BES_metro.m|  defines the states  given in Eq.~(\ref{eq:rho}). We also included other routines that show their usage. They are called 
\verb|example_BES_private.m|, 
\verb|example_BES_metro4x4.m|, and
\verb|example_BES_metro.m|.
These examples make it possible to verify that the quantum states have the metrological properties discussed in this paper. The programs \verb|BES_private.m| and \verb|BES_metro.m| can give the states corresponding to the order of the subsystems given as $ABA'B',$  as in this paper. The programs can also give the states corresponding to the order of the subsystems given as $AA'BB',$ which is more appropriate for studying bipartite entanglement between $AA'$ and $BB'.$


\begin{thebibliography}{10}
\bibitem{comment} The Hamiltonian for the numerically found states is of the form $D\otimes \openone+ \openone \otimes D$ where $D={\rm diag}(+1,+1,...,-1,-1),$ where there are $d$ $+1$'s in the diagonal, then $d$ $-1$'s. This is not the same as Eq.~(\ref{eq:Ham}), but can be obtained from it after trivial reordering of the parties to $AA'BB'.$ See also Appendix~\ref{sec:appendix_bound_private}.
\end{thebibliography}
\end{document}